\documentclass[journal]{IEEEtran}

\usepackage{epsfig,amsmath,amssymb,epsf,amsthm,scalefnt,multirow}
\usepackage{xcolor}
\usepackage{float}
\usepackage{cite}
\usepackage{psfrag}
\usepackage{gensymb}
\usepackage{cleveref}
\usepackage{mathrsfs}
\usepackage{mathtools}
\usepackage{algpseudocode}
\usepackage{algorithm}
\usepackage{enumerate}
\usepackage{stfloats}
\usepackage{latexsym}
\usepackage{multicol}
\usepackage[caption=false]{subfig}

\newtheorem{lemma}{Lemma}
\newtheorem*{lemma*}{Lemma}

  \def\cC{{\mathcal{C}}}

 \def\cN{{\mathcal{N}}} \def\cO{{\mathcal{O}}}

\def\diag{\mathop{\mathrm{diag}}}

\def\bSigma{{\pmb{\Sigma}}} 
\def\bPhi{{\pmb{\Phi}}} \def\bphi{{\pmb{\phi}}}

\def\bTheta{{\pmb{\Theta}}}

\def\b0{{\pmb{0}}} 

\def\ba{{\mathbf{a}}}   
 \def\bff{{\mathbf{f}}}  \def\bh{{\mathbf{h}}}
   
\def\bm{{\mathbf{m}}}   \def\bp{{\mathbf{p}}}
   
 \def\bv{{\mathbf{v}}}  \def\bx{{\mathbf{x}}}

\def\bA{{\mathbf{A}}}   \def\bD{{\mathbf{D}}}
\def\bE{{\mathbf{E}}} \def\bF{{\mathbf{F}}}  \def\bH{{\mathbf{H}}}
\def\bI{{\mathbf{I}}}   
   \def\bP{{\mathbf{P}}}
 \def\bR{{\mathbf{R}}}  
\def\bU{{\mathbf{U}}} \def\bV{{\mathbf{V}}} \def\bW{{\mathbf{W}}}

\begin{document}

\title{Multi-Group Multicasting Systems \\ Using Multiple RISs}

\author{Hyeongtaek~Lee, \textit{Member,~IEEE},
	Seungsik~Moon,
	Youngjoo~Lee, \textit{Senior Member, IEEE},
	Jaeky~Oh,
	Jaehoon~Chung,
	and~Junil~Choi, \textit{Senior Member, IEEE} 

	\thanks{Hyeongtaek~Lee and Junil~Choi are with the School of Electrical Engineering, Korea Advanced Institute of Science and Technology (e-mail: \{htlee8459; junil\}@kaist.ac.kr).}
	\thanks{Seungsik~Moon and Youngjoo~Lee are with the Department of Electrical Engineering, Pohang University of Science and Technology (e-mail: \{mssik1213;youngjoo.lee\}@postech.ac.kr).}
	\thanks{Jaeky~Oh and Jaehoon~Chung are with 6G Core R\&D Technology Platform (TP), Communication \& Media Standard Lab., ICT Tech. center, CTO Division, LG Electronics, Inc. (e-mail: \{jaeky.oh; jaehoon.chung\}@lge.com).}
	\thanks{This work was supported by LG Electronics Inc.; in part by the Ministry of Science and ICT (MSIT), South Korea, under the Information Technology Research Center (ITRC) Support Program supervised by the Institute of Information and Communications Technology Planning and Evaluation (IITP) under Grant IITP-2020-0-01787; and in part by the Korea Institute for Advancement of Technology (KIAT) grant funded by the Ministry of Trade, Industry and Energy (MOTIE) (P0022557).}
}
\maketitle

\begin{abstract}
	In this paper, practical utilization of multiple distributed reconfigurable intelligent surfaces (RISs), which are able to conduct group-specific operations, for multi-group multicasting systems is investigated. To tackle the inter-group interference issue in the multi-group multicasting systems, the block diagonalization (BD)-based beamforming is considered first. Without any inter-group interference after the BD operation, the multiple distributed RISs are operated to maximize the minimum rate for each group. Since the computational complexity of the BD-based beamforming can be too high, a multicasting tailored zero-forcing (MTZF) beamforming technique is proposed to efficiently suppress the inter-group interference, and the novel design for the multiple RISs that makes up for the inevitable loss of MTZF beamforming is also described. Effective closed-form solutions for the loss minimizing RIS operations are obtained with basic linear operations, making the proposed MTZF beamforming-based RIS design highly practical. Numerical results show that the BD-based approach has ability to achieve high sum-rate, but it is useful only when the base station deploys large antenna arrays. Even with the small number of antennas, the MTZF beamforming-based approach outperforms the other schemes in terms of the sum-rate while the technique requires low computational complexity. The results also prove that the proposed techniques can work with the minimum rate requirement for each group.
\end{abstract}

\section{Introduction} \label{Introduction}

Reconfigurable intelligent surface (RIS) has gained much interest to meet strong performance requirements of 6G and beyond wireless communication systems \cite{RIS_intro_1,RIS_intro_2,RIS_intro_3,RIS_intro_4,RIS_intro_5}. The RIS consists of a large number of low-cost passive elements that can induce desired phase shifts to impinging signals. When the phase shifts are carefully designed, the RIS is able to construct favorable wireless channel conditions and provide performance improvements, e.g., throughput enhancement and increased energy efficiency.

Recently, several works tried to exploit the RIS to assist multicasting systems. Note that a large portion of wireless data traffic is for common interest, for example, live broadcasting of sports games, popular videos, and news headlines \cite{Multicasting_intro_1}. Since these kinds of data can be effectively delivered by physical layer multicasting, the 3rd generation partnership project (3GPP) incorporated the evolved multimedia broadcast and multicast service (eMBMS) \cite{eMBMS_1,eMBMS_2}, and many researches have been conducted in recent years \cite{Multicasting_intro_1,Multicasting_intro_2,Multicasting_intro_3}. Different from unicast data transmissions where each user receives independent data, multiple users receive the same data in multicasting systems. Then, the multicasting systems can effectively alleviate the burden of high data traffic with reduced transmission delay, energy consumption, and hardware complexity \cite{Multicasting_intro_4,RIS_multicasting_4}.

For the single-group multicasting transmissions assisted by the RIS, the channel capacity was analyzed in \cite{RIS_single_group_multicasting_1}, an energy efficiency maximization problem was studied in \cite{RIS_single_group_multicasting_3}, the secrecy capacity was maximized in \cite{RIS_single_group_multicasting_4}, and the power minimization problem was considered with the rank-two Alamouti coding in \cite{RIS_single_group_multicasting_5}. These works, however, are based on the assumption of perfect channel state information (CSI). To alleviate the burden of CSI acquisition, adopting random RIS reflection coefficients was considered in~\cite{RIS_single_group_multicasting_2}. When adopting the RIS for multi-group multicasting systems, the transmit power was minimized considering quality-of-service (QoS) constraints in \cite{RIS_multicasting_4} and secure communications in \cite{RIS_multicasting_3}. In \cite{RIS_multicasting_1(single)}, an alternating optimization (AO) method was developed to maximize the sum-rate of all multicasting groups. With the rate-splitting multiple access scheme, the max-min fairness problem was considered in \cite{RIS_multicasting_5}. To deal with imperfect CSI issue, a robust design is discussed under the bounded CSI error model and the statistical CSI error model in \cite{RIS_multicasting_6}. Although these works made considerable performance improvements, they are limited to single-RIS-assisted systems. While \cite{Multi_RIS_1,Multi_RIS_3,Multi_RIS_4,Multi_RIS_6,Multi_RIS_2,Multi_RIS_5} considered multiple RISs, they are restricted to unicast transmission systems, and direct expansion to the multi-group multicasting systems is not possible. To the best of our knowledge, there has been no prior work that investigated the advantages of deploying multiple distributed RISs with their group-specific operations for the downlink multi-group multicasting systems.

To tackle this interesting scenario for the first time, we consider a downlink multi-group multicasting system exploiting multiple distributed RISs in this paper. A base station (BS) transmits independent data streams to each multicasting group while the users in the same group receive the same data. This implies there is no intra-group interference; however, there still exists inter-group interference that should be mitigated through proper beamforming at the BS. To resolve this issue, we propose the block-diagonalization (BD)-based and the multicasting tailored zero-forcing (MTZF) beamforming-based approaches. The main contributions of this paper are summarized as follows.

\begin{itemize}
	\item Following \cite{multicasting_block_diag}, we first consider the BD-based beamforming to completely eliminate the inter-group interference by exploiting null-space projections. Without any inter-group interference, each RIS is designed to maximize the minimum data rate of each group users. By applying the semi-definite relaxation (SDR) method, it is possible to obtain a fine approximate solution and properly handle the non-convex unit-norm constraint for each element of the RISs. Then, the beamformer at the BS and reflection coefficients of the multiple RISs are updated iteratively using the AO method. 
	\item Because the BD-based beamforming suffers from a large computational complexity, we also develop the MTZF beamforming technique at the BS to efficiently suppress the inter-group interference. However, the loss of intended signal power by nulling out the inter-group interference is inevitable for the MTZF beamforming. We adopt the Neumann series (NS) expansion method to represent the loss at each group, which can be minimized through the proposed closed-form solution of multiple RISs reflection coefficients. Since the solution does not require the exact value of MTZF beamformer, there is no need to alternately update the MTZF beamformer and multiple RISs reflection coefficients, which makes the computational complexity significantly lower than most of RIS designs that rely on the AO method including the proposed BD-based approach. In addition, the proposed MTZF beamforming and multiple RISs design only require basic linear operations, making the proposed technique highly~practical.
	\item Numerical results demonstrate that the BD-based approach can achieve high sum-rate when the BS deploys a large dimensional antenna array. When the number of BS antennas is comparable to the number of users, the BD-based approach cannot work well, while the MTZF beamforming-based approach shows the considerable performance improvement in terms of the sum-rate even with the notably low computational complexity. The results also show that the proposed techniques work under QoS constraints, e.g., the minimum rate requirement at each~group.
\end{itemize}

The remainder of this paper is organized as follows. In Section \ref{System Model}, we explain the system model of downlink multi-group multicasting system with multiple distributed RISs. The BD-based beamforming technique with the multiple RISs design to maximize the minimum data rate of each group is proposed in Section \ref{proposed BD+min. rate maximization RIS}. In Section \ref{proposed MTZF+loss minimization RIS}, the MTZF beamforming technique and the design for multiple RISs that minimize the total power of intended signal loss at each group is developed. Numerical results are shown in Section \ref{numerical results} to evaluate the performance of the proposed techniques, and the conclusion follows in Section \ref{conclusion}.

\textit{Notations}: Lower and upper boldface letters denote column vectors and matrices. The transpose and conjugate transpose of a matrix $\bA$ are represented by $\bA^\mathrm{T}$ and $\bA^\mathrm{H}$. For a square matrix $\bA$, $\bA^{-1}$ and $\mathrm{Tr}(\bA)$ are the inverse matrix and trace of $\bA$. The expression $\diag(\ba)$ is the diagonal matrix with the elements of $\ba$, and also $\diag(\bA)$ returns a diagonal matrix with the elements of $\bA$ on its main diagonal. $\angle(\ba)$ is the vector whose elements are the phase of each element of $\ba$. Notation $\vert a \vert$ is used for the absolute value of complex scalar number $a$. For a matrix $\bA$, $\bA(m,n)$ is the element of $m$-th row and $n$-th column, and $\ba(m)$ is the $m$-th element of a vector $\ba$. $\cC\cN(\mu,\sigma^2)$ stands for the complex Gaussian distribution with mean $\mu$ and variance $\sigma^2$. $\boldsymbol{0}_{m\times n}$ represents the $m \times n$ all-zero matrix, and $\bI_m$ is the $m \times m$ identity matrix. The $\ell_p$ norm of a vector $\ba$ is denoted by $\Vert \ba \Vert_p$. For a matrix $\bA$, $\Vert \bA \Vert$ denotes the spectral norm of $\bA$, and $\mathrm{rank}(\bA)$ implies the rank of $\bA$. $\cO(\cdot)$ implies the big-O notation. The Kronecker product is denoted~by $\otimes$.

\begin{figure}
	\centering
	\includegraphics[width=1.05\columnwidth]{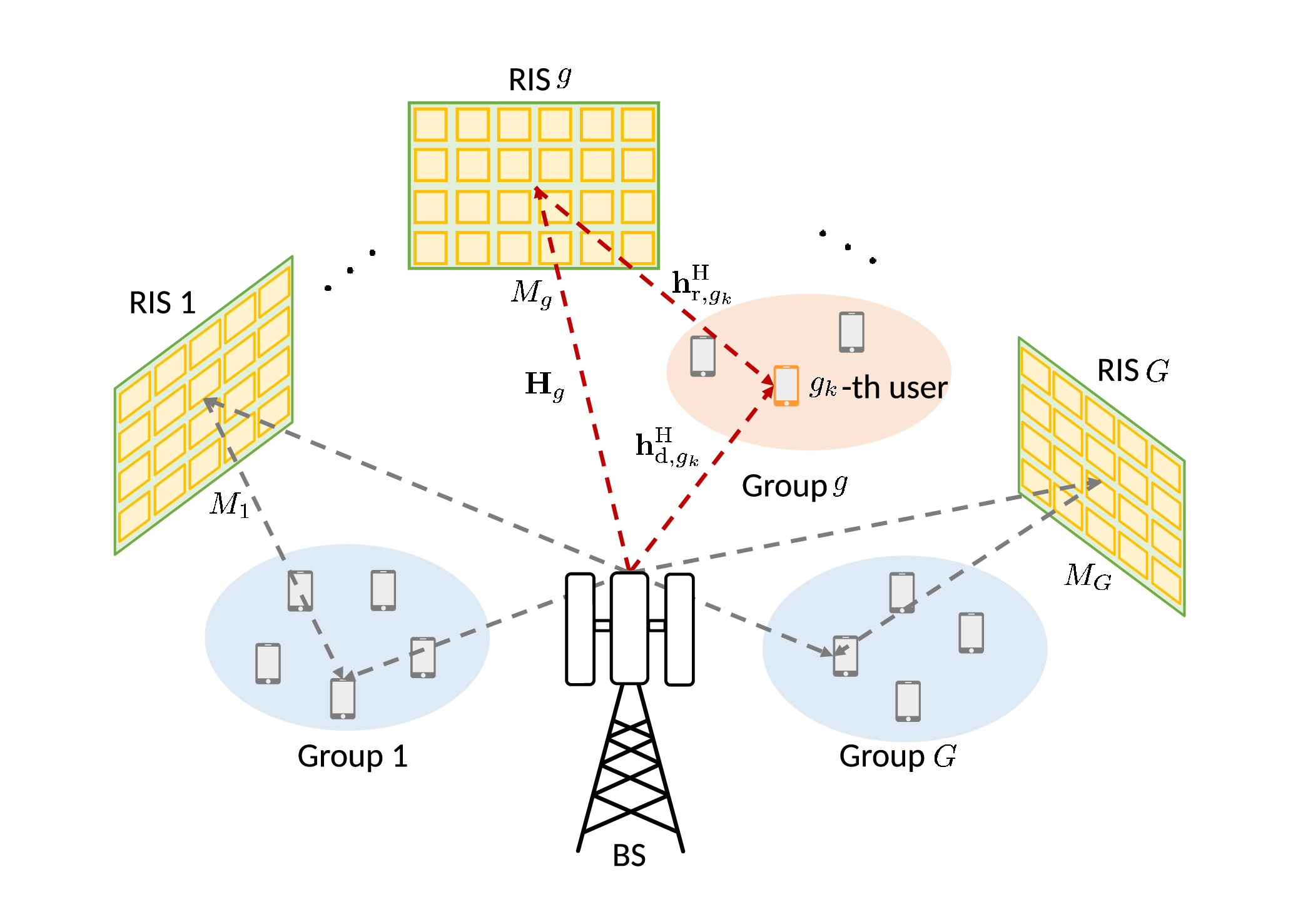}
	\caption{Conceptual figure of multi-group multicasting communication system utilizing  multiple distributed RISs.}
	\label{system model fig}
\end{figure}

\section{System Model} \label{System Model}

We consider a downlink multi-group multicasting communication system exploiting multiple distributed RISs as shown in Fig. \ref{system model fig}. The BS equipped with $N$ antennas serves $K$ single-antenna users divided into $G$ multicasting groups by their locations.\footnote{While other grouping strategies, e.g., using the second order statistics of their channels as in \cite{JSDM,JSDM_with_RIS}, are possible, we assume the users are grouped with their locations for simplicity.} Close to the $g$-th group users, an RIS, connected via a control link to the BS, with $M_g$ passive elements is placed to assist the users in group $g$. Let $K_g$ be the number of users in the $g$-th group such that $\sum_{g=1}^G K_g =K$. As in \cite{JSDM}, we define the index $g_k=\sum_{g'=1}^{g-1}K_{g'}+k$ to denote the $k$-th user in group $g$. 

Following the nature of multi-group multicasting systems, the BS transmits the same data to the users in the same group while independent data are transmitted to the different groups. Then, the downlink transmit signal $\bx$ at the BS is given~by
\begin{align}
	\bx=\sum_{g=1}^{G} \bff_g s_g,
\end{align}
where $\bff_g \in \mathbb{C}^{N \times 1}$ is the beamforming vector for the $g$-th group, and $s_g\in \mathbb{C}$ is the corresponding information data following $\mathbb{E}\left[\vert s_g \vert^2\right]=1$. We denote $\bF=[\bff_1,\cdots,\bff_G] \in \mathbb{C}^{N\times G}$ as the beamforming matrix at the BS satisfying the total power constraint $\mathrm{Tr}\left( \bF^\mathrm{H} \bF\right) \leq P_\mathrm{T}$ with the maximum downlink transmit power $P_\mathrm{T}$.

For the users in the $g$-th group, we assume the signals reflected through the other RISs except the $g$-th RIS are negligible because of large path loss \cite{Multi_RIS_1,Multi_RIS_3,Multi_RIS_4,Multi_RIS_6}. Then, the downlink received signal at the $g_k$-th user is
\begin{align}
	y_{g_k}=\left(\bh_{\mathrm{d},g_k}^\mathrm{H}+\bh_{\mathrm{r},g_k}^\mathrm{H}\bPhi_g \bH_g \right)\sum_{g'=1}^{G}\bff_{g'} s_{g'}+n_{g_k}, \label{downlink received signal model}
\end{align}
where $n_{g_k}\sim \cC\cN\left(0,\sigma_{g_k}^2\right)$ is the additive thermal noise. The reflection coefficient matrix at the $g$-th RIS is defined by the $M_g \times M_g$ diagonal matrix $\bPhi_g \triangleq \diag(\bphi_g)$ with the reflection coefficient vector $\bphi_g \in \mathbb{C}^{M_g \times 1}$ where $\vert\bphi_g(m_g)\vert=1$ for $m_g=1,\cdots,M_g$. The BS-user direct link, RIS-user reflection link, and BS-RIS link are denoted by $\bh_{\mathrm{d},g_k}^\mathrm{H} \in \mathbb{C}^{1 \times N}$, $\bh_{\mathrm{r},g_k}^\mathrm{H} \in \mathbb{C}^{1 \times M_g}$, and $\bH_g \in \mathbb{C}^{M_g \times N}$, respectively.\footnote{Since the proposed technique does not rely on a specific channel model, we intentionally do not assume any model here.} Since the channels for all links including the RISs can be estimated by the existing channel estimation techniques \cite{RIS_channel_est_1,RIS_channel_est_2,Min_rate_max_RIS}, we assume the perfect CSI at the BS to handle the multi-group multicasting systems deploying multiple RISs for the first time.

With the received signal model in \eqref{downlink received signal model}, the data rate of the $g_k$-th user $R_{g_k}$ can be obtained by \cite{RIS_multicasting_1(single)}
\begin{align}
	R_{g_k}=\log_2 \left(1+\frac{\left\vert \bh_{g_k}^\mathrm{H}\bff_g \right\vert^2}{\sum_{{g'}\neq g}^{G} \left\vert \bh_{g_k}^\mathrm{H}\bff_{g'} \right\vert^2+\sigma_{g_k}^2} \right),
	\label{rate of g_k user}
\end{align}
where we denote the total downlink channel of the $g_k$-th user as $\bh_{g_k}^\mathrm{H}=\bh_{\mathrm{d},g_k}^\mathrm{H}+\bh_{\mathrm{r},g_k}^\mathrm{H}\bPhi_g \bH_g$. Considering the inherent characteristic of multicasting systems that the users in the same group receive the same data, the data rate of each group is bounded by the minimum rate of users in the group. Therefore, the sum-rate of all groups $R_\mathrm{sum}$ is given by 
\begin{align}
	R_\mathrm{sum}=\sum_{g=1}^{G} \enspace \min_{g_k \in \{g_1,\cdots,g_{K_g} \}} R_{g_k}.
	\label{multicasting sum rate}
\end{align}
While there is no intra-group interference in the multi-group multicasting systems, the inter-group interference must be properly handled to improve the data rate of each user. In the following Sections \ref{proposed BD+min. rate maximization RIS} and \ref{proposed MTZF+loss minimization RIS}, we describe the two proposed techniques that mitigate the inter-group interference, and the multiple RISs are designed appropriately to each beamforming technique at the BS.


\section{BD-based Beamforming \\ with Minimum Rate Maximizing RISs Design} \label{proposed BD+min. rate maximization RIS}

For the typical multi-user multiple-input multiple-output (MU-MIMO) unicast transmission systems, the BD-based beamforming technique is developed to completely eliminate the interference among the users \cite{conventional_block_diag}, and the principal idea is adopted for the multi-group multicasting systems to mitigate the inter-group interference in \cite{multicasting_block_diag}. In Section \ref{BD-based beamforming technique}, we first explain the BD-based beamforming technique developed in \cite{multicasting_block_diag} for completeness, and the multiple RISs design to maximize the minimum rate of users for each group is elaborated in Section \ref{Minimum rate maximizing RISs design}. We summarize the proposed technique and analyze its complexity in Section \ref{BD-based approach complexity}.

\subsection{BD-Based Beamforming Technique} \label{BD-based beamforming technique}

We assume the reflection coefficient matrices of multiple RISs are fixed and define $\bH_{\mathrm{tot},g} \in \mathbb{C}^{K_g \times N}$ and $\check{\bH}_{\mathrm{tot},g} \in \mathbb{C}^{(K-K_g) \times N}$ as the channel matrix of all users belonging to group $g$ and all users not belonging to group $g$, respectively, which are given by
\begin{align}
	\bH_{\mathrm{tot},g} &\triangleq \left[\bh_{g_1}, \cdots, \bh_{g_{K_g}}\right]^\mathrm{H}, \\
	\check{\bH}_{\mathrm{tot},g} &\triangleq \left[\bH_{\mathrm{tot},1}^\mathrm{H},\cdots,\bH_{\mathrm{tot},g-1}^\mathrm{H},\bH_{\mathrm{tot},g+1}^\mathrm{H},\cdots,\bH_{\mathrm{tot},G}^\mathrm{H} \right]^\mathrm{H}.
\end{align}
By applying the singular value decomposition (SVD), $\check{\bH}_{\mathrm{tot},g}$ can be represented as follows
\begin{align}
	\check{\bH}_{\mathrm{tot},g} &= \check{\bU}_g \check{\bSigma}_g \check{\bV}_g^\mathrm{H}, \notag \\
	&=\check{\bU}_g \check{\bSigma}_g \left[\check{\bV}_g^{(1)}, \check{\bV}_g^{(0)} \right]^\mathrm{H},
\end{align}
where $\check{\bU}_g \in \mathbb{C}^{(K-K_g) \times (K-K_g)}$ is a unitary matrix and $\check{\bSigma}_g \in \mathbb{C}^{(K-K_g) \times N}$ contains the singular values on its diagonal. The unitary matrix $\check{\bV}_g \in \mathbb{C}^{N \times N}$ can be divided into $\check{\bV}_g^{(1)} \in \mathbb{C}^{N \times \check{r}_g}$ and $\check{\bV}_g^{(0)} \in \mathbb{C}^{N \times (N-\check{r}_g)}$ where $\check{r}_g=\mathrm{rank}\left(\check{\bH}_{\mathrm{tot},g}\right)$. Then, the column vectors of $\check{\bV}_g^{(0)}$ become the right singular vectors of $\check{\bH}_{\mathrm{tot},g}$ corresponding to the zero singular values. Therefore, the column vectors constitute an orthonormal basis for the null-space of $\check{\bH}_{\mathrm{tot},g}$, i.e., $\bH_{\mathrm{tot},g'}\check{\bV}_g^{(0)}=\boldsymbol{0}_{K_{g'} \times (N-\check{r}_g)}$ for $g' \neq g$.

Since data can be transmitted to the $g$-th group only when the null-space of $\check{\bH}_{\mathrm{tot},g}$ has a dimension greater than zero, we have the necessary condition $N > \check{r}_g$, and this condition limits the number of $K$ that can be supported simultaneously.\footnote{Although the condition is satisfied, as shown in Section \ref{numerical results}, the performance of BD-based beamforming is not sufficient if $N$ is comparable to $K$. This is because, when the BS has only a moderate number of antennas, the technique suffers from severe intended signal power loss from the interference cancellation considering all channels in the system.} When the condition is satisfied, $\check{\bV}_g^{(0)}$ can be used to construct the beamforming vector for the $g$-th group that eliminates the interference from the other groups. We denote $\bH_{\mathrm{tot},g}^{\mathrm{eq}}=\bH_{\mathrm{tot},g} \check{\bV}_g^{(0)} \in \mathbb{C}^{K_g \times (N-\check{r}_g)}$ as the equivalent channel matrix of group $g$ after the null-space projection and construct the equivalent beamforming vector $\bm_{g}^\mathrm{eq}\in\mathbb{C}^{(N-\check{r}_g)\times 1}$ as the maximum eigenvector of $\left(\bH_{\mathrm{tot},g}^{\mathrm{eq}}\right)^\mathrm{H} \bH_{\mathrm{tot},g}^{\mathrm{eq}}$ as in \cite{multicasting_block_diag}. Then, the final beamforming matrix by the BD-based beamforming technique $\bF_\mathrm{BD}$ is given~by
\begin{align}
\bF_\mathrm{BD} &= \left[\bff_{\mathrm{BD},1},\cdots,\bff_{\mathrm{BD},G}\right], \notag \\
&= \left[\check{\bV}_1^{(0)}\bm_{1}^\mathrm{eq},\cdots,\check{\bV}_G^{(0)}\bm_{G}^\mathrm{eq} \right] \bP_\mathrm{BD}^{1/2}, \label{proposed solution of BD-based beamforming}
\end{align} 
where $\bP_\mathrm{BD}=\diag\left([p_{\mathrm{BD},1},\cdots,p_{\mathrm{BD},G}]^\mathrm{T}\right)\in \mathbb{C}^{G \times G}$ is the diagonal power allocation matrix. While other advanced power allocation methods are also possible, we adopt the equal power allocation for $\bP_\mathrm{BD}$ for simplicity in this paper.

\subsection{Minimum Rate Maximizing Multiple RISs Design} \label{Minimum rate maximizing RISs design}

With given $\bF_\mathrm{BD}$ in \eqref{proposed solution of BD-based beamforming}, it is possible to process each group individually, and $R_{g_k}$ without any inter-group interference simply becomes
\begin{align}
	R_{g_k}=\log_2 \left(1+\frac{\left\vert \bh_{g_k}^\mathrm{H}\bff_{\mathrm{BD},g} \right\vert^2}{\sigma_{g_k}^2}  \right).
\end{align} 
Following the multicasting nature, as we discussed in Section~\ref{System Model}, the data rate of group $g$ is given by the minimum rate $\min_{g_k \in \{g_1,\cdots,g_{K_g} \}} R_{g_k}$. Consequently, we can design $\bPhi_g$ to consider the fairness problem, i.e., maximize the minimum rate. Assuming the other group's RIS reflection coefficient matrices are fixed, the optimization problem $(\mathrm{P}1)$ for $\bPhi_g$ is given~as
\begin{align}
	(\mathrm{P}1): &\max_{\bPhi_g} \min_{g_k \in \{g_1,\cdots,g_{K_g} \}} \enspace \frac{\left\vert \bh_{g_k}^\mathrm{H}\bff_{\mathrm{BD},g} \right\vert^2}{\sigma_{g_k}^2} \notag \\
	&\enspace\mathrm{s.t.} \enspace \bPhi_g=\diag(\bphi_g), \notag \\
	& \quad\enspace\enspace \vert\bphi_g(m_g)\vert=1, \enspace m_g=1,\cdots,M_g. \label{min rate maximization problem}
\end{align}
Although the objective function is affine, $(\mathrm{P}1)$ is a non-convex problem due to the unit-norm constraints on $\bphi_g$. To get a solution of $(\mathrm{P}1)$, we represent the objective function in \eqref{min rate maximization problem} into an equivalent quadratic form as 
\begin{align}
	&\frac{\left\vert \bh_{g_k}^\mathrm{H}\bff_{\mathrm{BD},g} \right\vert^2}{\sigma_{g_k}^2} \notag \\ 
	=&\frac{1}{\sigma_{g_k}^2}\left\vert \left(\bh_{\mathrm{d},g_k}^\mathrm{H}+\bh_{\mathrm{r},g_k}^\mathrm{H}\bPhi_g \bH_g \right) \bff_{\mathrm{BD},g}\right\vert^2, \notag \\
	\stackrel{(\mathrm{a})}{=}&\frac{1}{\sigma_{g_k}^2}\left\vert \left(\bh_{\mathrm{d},g_k}^\mathrm{H}+\bphi_g^\mathrm{H}\diag\left(\bh_{\mathrm{r},g_k}\right) \bH_g \right) \bff_{\mathrm{BD},g}\right\vert^2, \notag \\
	=& \frac{1}{\sigma_{g_k}^2}\Big\vert \underbrace{\bh_{\mathrm{d},g_k}^\mathrm{H}\bff_{\mathrm{BD},g}}_{\triangleq q_{g_k}}+\bphi_g^\mathrm{H}\underbrace{\diag\left(\bh_{\mathrm{r},g_k}\right) \bH_g\bff_{\mathrm{BD},g}}_{\triangleq \bp_{g_k}} \Big\vert^2, \notag \\
	=&\frac{1}{\sigma_{g_k}^2}\left(\bphi_g^\mathrm{H}\bp_{g_k}\bp_{g_k}^\mathrm{H}\bphi_g+\bphi_g^\mathrm{H}\bp_{g_k}q_{g_k}^*+ q_{g_k}\bp_{g_k}^\mathrm{H}\bphi_g+ \left\vert q_{g_k} \right\vert^2\right), \label{min rate maximization problem ojbective function quadratic}
\end{align}
where $(\mathrm{a})$ can be derived by using the definition $\bPhi_g \triangleq \diag\left( \bphi_g \right)$. With an auxiliary variable $t_g$ and the quadratic form in \eqref{min rate maximization problem ojbective function quadratic}, $(\mathrm{P}1)$ is given as 
\begin{align}
	(\mathrm{P}1): &\max_{\bar{\bphi}_g} \min_{g_k \in \{g_1,\cdots,g_{K_g} \}} \enspace \bar{\bphi}_g^\mathrm{H} \bW_{g_k} \bar{\bphi}_g + \frac{\left\vert q_{g_k} \right\vert^2}{\sigma_{g_k}^2} \notag \\
	&\enspace\mathrm{s.t.} \enspace \left\vert \bar{\bphi}_g (\bar{m}_g) \right\vert =1, \enspace \bar{m}_g=1,\cdots,M_g+1, 
\end{align}
where 
\begin{align}
\bW_{g_k}=\frac{1}{\sigma_{g_k}^2}
\begin{bmatrix} 
	\bp_{g_k}\bp_{g_k}^\mathrm{H} & \bp_{g_k}q_{g_k}^* \\ q_{g_k}\bp_{g_k}^\mathrm{H} & 0 
\end{bmatrix},\enspace 
\bar{\bphi}_g=
\begin{bmatrix}
	\bphi_g \\ t_g
\end{bmatrix}. \notag
\end{align}

Note that $\bar{\bphi}_g^\mathrm{H} \bW_{g_k} \bar{\bphi}_g = \mathrm{Tr}\left(\bW_{g_k}\bar{\bphi}_g\bar{\bphi}_g^\mathrm{H}\right)=\mathrm{Tr}\left(\bW_{g_k}\bar{\bTheta}_g\right)$ by defining $\bar{\bTheta}_g \triangleq \bar{\bphi}_g\bar{\bphi}_g^\mathrm{H}$, which needs to satisfy $\bar{\bTheta}_g \succeq 0$ and $\mathrm{rank}\left(\bar{\bTheta}_g\right)=1$. Then, we adopt the SDR method as in \cite{RIS_intro_5,Min_rate_max_RIS} to relax the non-convex rank-one constraint, and let $\bar{\bTheta}_g$ be a positive semi-definite matrix of arbitrary rank. The resulting problem $(\mathrm{P}2)$ is reduced to
\begin{align}
	(\mathrm{P}2): &\max_{\bar{\bTheta}_g} \min_{g_k \in \{g_1,\cdots,g_{K_g} \}} \enspace \mathrm{Tr}\left(\bW_{g_k}\bar{\bTheta}_g \right) + \frac{\left\vert q_{g_k} \right\vert^2}{\sigma_{g_k}^2} \notag \\
	&\enspace\mathrm{s.t.} \enspace \bar{\bTheta}_g \succeq 0, \notag \\
	& \quad\enspace\enspace \left\vert \bar{\bTheta}_g (\bar{m}_g,\bar{m}_g) \right\vert =1, \enspace \bar{m}_g=1,\cdots,M_g+1, 
\end{align}
which is a convex optimization problem that can be solved optimally by existing solvers such as CVX \cite{cvx}. In general, however, the optimal solution of $(\mathrm{P}2)$ may not satisfy the rank-one constraint, i.e., $\mathrm{rank}\left(\bar{\bTheta}_g \right)\neq 1$. Therefore, the Gaussian randomization method as in \cite{Min_rate_max_RIS} will be additionally required to get a rank-one solution from the obtained higher-rank solution of $(\mathrm{P}2)$ where sufficiently many randomization guarantees very tight approximation of the objective value of $(\mathrm{P}2)$~\cite{SDR}. With the final $\bar{\bphi}_g^\star$ after the Gaussian randomization, the effective reflection coefficient vector $\bphi_{\mathrm{BD},g}^\star$ can be obtained~as
\begin{align}
	\bphi_{\mathrm{BD},g}^\star=\mathrm{exp}\left(j\angle\left(\left[\frac{\bar{\bphi}_g^\star}{\bar{\bphi}_g^\star(M_g+1)}\right]_{(1:M_g)}\right)\right), \label{proposed solution of min rate max RIS}
\end{align}
where $\left[ \ba\right]_{(1:m)}$ denotes the vector of first $m$ elements of $\ba$.

\alglanguage{pseudocode}
\begin{algorithm}[t]
	\caption{Proposed BD-based beamforming with minimum rate maximizing multiple RISs design}
	\textbf{Initialize}
	\begin{algorithmic}[1]
		\State Set $\epsilon_{\mathrm{rate}} >0$ and $R_\mathrm{tmp}=0$
		\State Initialize $\{\bphi_{\mathrm{BD},g}\}_{g=1}^G$
	\end{algorithmic}
	\textbf{Iterative update}
	\begin{algorithmic}[1]
		\addtocounter{ALG@line}{+2}
		\For{$i_1=1,\cdots,I_1$}
		\State Compute $\bF_\mathrm{BD}$ by \eqref{proposed solution of BD-based beamforming}
		\For{$g=1,\cdots,G$}
		\State Update $\bphi_{\mathrm{BD},g}^\star$ by \eqref{proposed solution of min rate max RIS}
		\EndFor
		\State Calculate $R_\mathrm{sum}$ by \eqref{multicasting sum rate}
		\If {$\vert R_\mathrm{sum}-R_\mathrm{tmp}\vert < \epsilon_{\mathrm{rate}}$}
		\State Break
		\Else
		\State Set $R_\mathrm{tmp}=R_\mathrm{sum}$
		\EndIf
		\EndFor
		\State Update $\bF_\mathrm{BD}$ by \eqref{proposed solution of BD-based beamforming} with final $\left\{\bphi_{\mathrm{BD},g}^\star\right\}_{g=1}^G$
	\end{algorithmic}
\end{algorithm}

\subsection{Algorithm Description and Complexity Analysis} \label{BD-based approach complexity}

The overall process of the proposed BD-based beamforming and the minimum rate maximizing multiple RISs design is summarized in Algorithm 1. By exploiting the AO method, the beamforming matrix $\bF_\mathrm{BD}$ at the BS and the reflection coefficient vectors $\left\{\bphi_{\mathrm{BD},g}^\star\right\}_{g=1}^G$ at the multiple RISs are updated in an iterative manner until when the iteration index $i_1$ reaches its maximum value $I_1$ or the sum-rate difference is less than the threshold $\epsilon_{\mathrm{rate}}$.\footnote{Although it is difficult to mathematically prove the convergence of Algorithm 1, we numerically verified that the sum-rate of BD-based approach converges within a few iterations.} Note that the last update of $\bF_\mathrm{BD}$ is conducted to eliminate the inter-group interference with the designed reflection coefficients of multiple RISs.

In order to analyze the computational complexity of Algorithm 1, we count the number of required complex-valued multiplications assuming $N \gg G$. For each iteration, the computational complexity of the BD-based beamforming is given by $\cO\big(5N^3 G-N^2KG+3NK^2G-K^3G\big)$ according to \cite{multicasting_block_diag}. Also, the update of the multiple RISs reflection coefficient vectors requires multiplications of order $\cO\big(\sum_{g=1}^G\big(2NM_g+M_g^{4.5}\big)\big)$, which is mainly dominated by obtaining the solution of $(\mathrm{P}2)$~\cite{SDR}. Then, the overall maximum complexity becomes $\cO\big(I_1\big(5N^3 G-N^2KG+3NK^2G-K^3G+\sum_{g=1}^G\big(2NM_g+M_g^{4.5}\big)\big)\big)$. Note that the proposed BD-based approach in this section can eliminate inter-group interference and maximize the minimum rate of each group effectively; however, it will suffer from the huge computational complexity when there are many users and the number of each RIS elements increases.
 
\section{MTZF Beamforming \\ with Loss Minimizing RISs Design} \label{proposed MTZF+loss minimization RIS}

Although effective, the BD-based beamforming requires high computational complexity due to the multiple large-dimensional SVD operations. To overcome the complexity issue, in this section, we also propose the MTZF beamforming to efficiently mitigate the inter-group interference for the multi-group multicasting systems. Then, for the reflection coefficients of multiple RISs, a novel design to minimize the total power of intended signal loss of each group, which is inevitable for the MTZF beamforming, is developed. It will be shown that, different from most of previous works and the BD-based approach proposed in Section \ref{proposed BD+min. rate maximization RIS}, there is no need to consider the AO method to optimize the MTZF beamforming at the BS and the multiple RISs reflection coefficients. Furthermore, the proposed design gives an effective closed-form solution where all the procedures can be conducted by basic linear operations with low complexity, which makes the proposed design highly practical.

\subsection{Proposed MTZF Beamforming Technique} \label{proposed MTZF beamforming}

In addition to the BD-based beamforming, the conventional linear ZF beamforming is widely exploited to mitigate the interference among the users in the MU-MIMO unicast transmission systems as in \cite{Linear_ZF_1,Linear_ZF_2} where the independent data is transmitted through the beamformer for each user. Unfortunately, for the multi-group multicasting systems, the linear ZF beamforming cannot be directly applied to suppress the inter-group interference since the users in the same group receive the same data through one group-specific beamforming vector even though their channels would not be the same in general. Therefore, we consider a simple yet effective approach of constructing representative channels (RCs) to exploit the linear ZF beamforming for the multi-group multicasting systems.

Considering physical geometry of the wireless channel environments, the channel characteristics of users in the same group will be similar since the channels are determined by their locations, scattering, antenna array at the BS, and element array at the RISs \cite{Multi_RIS_1,Geometric_channel}. This allows us to construct a RC for each group. Expecting the RCs would give an appropriate description of each group's channels, we can handle all channels of each group with their RC and exploit the RCs as sufficient information of whole user channels for the beamforming at the BS.

With the perfect CSI assumption, we can formulate the RCs by taking the arithmetic mean of each group user channels, and the RC of $g$-th group $\tilde{\bh}_g$ is given by

\begin{align}
	\tilde{\bh}_g^\mathrm{H}
	 & = \frac{1}{K_g} \sum_{k=1}^{K_g} \bh_{g_k}^\mathrm{H}, \notag                                                                                                   \\
	 & = \frac{1}{K_g} \sum_{k=1}^{K_g} \left(\bh_{\mathrm{d},g_k}^\mathrm{H}+\bh_{\mathrm{r},g_k}^\mathrm{H}\bPhi_g \bH_g \right), \notag                 \\
	 & = \frac{1}{K_g}  \left(\sum_{k=1}^{K_g}\bh_{\mathrm{d},g_k}^\mathrm{H}+\sum_{k=1}^{K_g}\bh_{\mathrm{r},g_k}^\mathrm{H}\bPhi_g \bH_g \right), \notag \\
	 & =\tilde{\bh}^\mathrm{H}_{\mathrm{d},g}+\tilde{\bh}^\mathrm{H}_{\mathrm{r},g}\bPhi_g \bH_g,  \label{rerpesentative channel definition}
\end{align}
where the RC of BS-user direct and RIS-user reflection links of the $g$-th group are defined by $\tilde{\bh}^\mathrm{H}_{\mathrm{d},g}\triangleq \frac{1}{K_g}\sum_{k=1}^{K_g}\bh_{\mathrm{d},g_k}^\mathrm{H}$ and $\tilde{\bh}^\mathrm{H}_{\mathrm{r},g}\triangleq \frac{1}{K_g}\sum_{k=1}^{K_g}\bh_{\mathrm{r},g_k}^\mathrm{H}$.

With the RC vectors of all groups, we can define the RC matrix $\tilde{\bH}=\left[\tilde{\bh}_1,\cdots,\tilde{\bh}_G \right]\in \mathbb{C}^{N \times G}$. Then, the proposed MTZF beamforming matrix is $\bF_{\mathrm{ZF}}$ given by
\begin{align}
	\bF_{\mathrm{ZF}} &= \left[\bff_{\mathrm{ZF},1},\cdots,\bff_{\mathrm{ZF},G}\right], \notag \\ &=\tilde{\bH} \left(\tilde{\bH}^\mathrm{H} \tilde{\bH} \right)^{-1} \bP_{\mathrm{ZF}}^{1/2}, \label{MTZF beamforming}
\end{align}
where $\bP_{\mathrm{ZF}}=\diag\left([p_{\mathrm{ZF},1},\cdots,p_{\mathrm{ZF},G}]^\mathrm{T}\right)\in \mathbb{C}^{G \times G}$ is the diagonal power allocation matrix. As in the previous section, the equal power allocation is adopted for $\bP_\mathrm{ZF}$ in this paper.

Even with the matrix inversion in~\eqref{MTZF beamforming}, the computational complexity of proposed MTZF beamforming is much lower than that of the BD-based beamforming in Section \ref{BD-based beamforming technique}. This is because the computational complexity of matrix inversion in \eqref{MTZF beamforming} only increases with the number of groups $G$ that would be much smaller than $N$ and $K$ in practice. In addition, the MTZF beamforming technique works only if the condition $N \geq G$ is satisfied regardless of the number of $K$ since the technique exploits the RC for each group.

We can anticipate the reduced inter-group interference by exploiting the MTZF beamforming in \eqref{MTZF beamforming}. Still, there is a weak point of the MTZF beamforming. When the information data $s_g$ is transmitted with its corresponding MTZF beamforming vector $\bff_{\mathrm{ZF},g}$, the intended signal power received at the $g_k$-th user is $\left\vert \bh_{g_k}^\mathrm{H} \bff_{\mathrm{ZF},g} \right\vert^2$ that is smaller than the case of maximum ratio transmission. Although inevitable for all ZF beamforming families, the amount of the loss can be adjusted by varying the channel conditions using the multiple RISs. In the following subsections, we develop the multiple RISs reflection coefficients to minimize the intended signal power loss of MTZF beamforming.

\subsection{Low-Complexity Approximation of MTZF Beamforming Vectors without Matrix Inversion}

In order to design the loss minimizing RIS reflection coefficients, the first thing to do is to quantify the loss at each user in terms of the channel vectors consisting of the reflection coefficients. To do this, the MTZF beamforming vectors $\left\{ \bff_{\mathrm{ZF},g} \right\}_{g=1}^G$ should be fully expressed with the channel vectors since the loss at each user will be represented with the channels and MTZF beamforming vectors. This is a difficult problem though due to the matrix inversion in \eqref{MTZF beamforming}. To resolve this issue, we adopt the NS expansion, which was originally developed to reduce the computational complexity of the inverse operation, to transform the matrix inversion with simple matrix multiplications and summations \cite{NS_1,NS_2,NS_3}.

After defining the Gram matrix $\bR \triangleq \left(\tilde{\bH}^\mathrm{H} \tilde{\bH} \right)/N$, the inverse of $\bR$ can be expressed through the NS expansion as~\cite{NS_2}
\begin{align}
	\bR^{-1} = \sum_{\ell=0}^{\infty} \left(-\bD^{-1}\bE \right)^\ell \bD^{-1},
	\label{NS expansion}
\end{align}
with the precondition matrix $\bD\in \mathbb{C}^{G \times G}$ and $\bE \triangleq\bR-\bD$. The equality in \eqref{NS expansion} holds if $\lim_{\ell \rightarrow \infty} \left(-\bD^{-1}\bE \right)^\ell = \boldsymbol{0}_{G\times G}$ or $\left\Vert -\bD^{-1}\bE \right\Vert<1$. To make things practical, we only take a finite summation in \eqref{NS expansion} as
\begin{align}
	\bR^{-1} \approx \sum_{\ell=0}^{L} \left(-\bD^{-1}\bE \right)^\ell \bD^{-1},
	\label{NS expansion - approx}
\end{align}
where large $L$ gives better approximation. Even with \eqref{NS expansion - approx}, it is quite complicated to express the effect of RCs on the MTZF beamformer. Therefore, we consider $L=1$ as in \cite{NS_2}, which gives the approximated MTZF beamforming matrix as
\begin{align}
	\bF_{\mathrm{ZF}}
	 & =\tilde{\bH}\left(\tilde{\bH}^\mathrm{H}\tilde{\bH}\right)^{-1} \bP_{\mathrm{ZF}}^{1/2}, \notag                      \\
	 & =\frac{1}{N} \tilde{\bH} \bR^{-1}\bP_{\mathrm{ZF}}^{1/2}, \notag                                                     \\
	 & \approx \frac{1}{N} \tilde{\bH} \left(\bD^{-1}-\bD^{-1} \bE \bD^{-1} \right)\bP_{\mathrm{ZF}}^{1/2},\notag           \\
	 & =\frac{1}{N} \tilde{\bH} \left(\bD^{-1}-\bD^{-1} \left(\bR-\bD\right) \bD^{-1} \right)\bP_{\mathrm{ZF}}^{1/2},\notag \\
	 & =\frac{1}{N}\tilde{\bH}\left(2 \bD^{-1}-\bD^{-1}\bR\bD^{-1} \right)\bP_{\mathrm{ZF}}^{1/2}, \notag \\ 
	 & =\frac{1}{N}\tilde{\bH}\bD^{-1}\left(2\bI_G-\bR\bD^{-1}\right)\bP_{\mathrm{ZF}}^{1/2}.
	\label{NS approximation - MTZF beamforming}
\end{align}

It is clear from \eqref{NS approximation - MTZF beamforming} that the choice of $\bD$ would significantly affect the approximation performance and complexity trade-off. Among possible candidates, we adopt the diagonal NS (DNS) method in \cite{NS_1} that showed satisfying approximation performance especially when the diagonal elements of $\bR$ are dominant. This is a valid assumption when the BS is deployed with a large number of antennas \cite{Channel_hardening}. Applying the DNS method, the diagonal precondition matrix $\bD$ is given~as
\begin{align}
	\bD
	&=\diag(\bR), \notag \\
	&=\frac{1}{N}
	\begin{bmatrix}
		\Vert \tilde{\bh}_1 \Vert_2^2 &        &                               \\
		                              & \ddots &                               \\
		                              &        & \Vert \tilde{\bh}_G \Vert_2^2
	\end{bmatrix}.
\end{align}
Then, the approximation of the MTZF beamforming matrix in~\eqref{NS approximation - MTZF beamforming} is represented by \eqref{NS approximation - MTZF beamforming matrix} in the top of next page,
\begin{figure*}[t]
\begin{align}
	\bF_{\mathrm{ZF}}= & \left[\tilde{\bh}_1,\cdots,\tilde{\bh}_G\right]
	\begin{bmatrix}
		\Vert \tilde{\bh}_1 \Vert_2^2 &        &                               \\
		                              & \ddots &                               \\
		                              &        & \Vert \tilde{\bh}_G \Vert_2^2
	\end{bmatrix}^{-1}
	\left(2\bI_G - 
	      \begin{bmatrix}
		       \Vert \tilde{\bh}_1 \Vert_2^2         & \cdots & \tilde{\bh}_1^\mathrm{H}\tilde{\bh}_G \\
		       \vdots                                & \ddots & \vdots                                \\
		       \tilde{\bh}_G^\mathrm{H}\tilde{\bh}_1 & \cdots & \Vert \tilde{\bh}_G \Vert_2^2
	       \end{bmatrix}
	\begin{bmatrix}
		\Vert \tilde{\bh}_1 \Vert_2^2 &        &                               \\
		                              & \ddots &                               \\
		                              &        & \Vert \tilde{\bh}_G \Vert_2^2
	\end{bmatrix}^{-1}\right) \notag \\
	& \times \begin{bmatrix}
		\sqrt{p_{\mathrm{ZF},1}} &        &            \\
		           & \ddots &            \\
		           &        & \sqrt{p_{\mathrm{ZF},G}}
	\end{bmatrix}, \notag \\
	= & \left[ \frac{\tilde{\bh}_1}{\Vert \tilde{\bh}_1 \Vert_2^2}, \cdots, \frac{\tilde{\bh}_G}{\Vert \tilde{\bh}_G \Vert_2^2} \right]
	\begin{bmatrix}
		1 & \cdots & -\frac{\tilde{\bh}_1^\mathrm{H}\tilde{\bh}_G}{\Vert \tilde{\bh}_G \Vert_2^2} \\
		\vdots & \ddots & \vdots \\
		-\frac{\tilde{\bh}_G^\mathrm{H}\tilde{\bh}_1}{\Vert \tilde{\bh}_1 \Vert_2^2} & \cdots & 1
	\end{bmatrix} 
	\begin{bmatrix}
		\sqrt{p_{\mathrm{ZF},1}} &        &            \\
		           & \ddots &            \\
		           &        & \sqrt{p_{\mathrm{ZF},G}}
	\end{bmatrix}
	\label{NS approximation - MTZF beamforming matrix}
\end{align}
\par\noindent\rule{\textwidth}{0.4pt}
\end{figure*}
which gives the approximated $g$-th beamforming vector $\bff_{{\mathrm{ZF}},g}$ as
\begin{align}
	\bff_{{\mathrm{ZF}},g}= \sqrt{p_{{\mathrm{ZF}},g}}\left(\underbrace{\frac{\tilde{\bh}_g}{\Vert \tilde{\bh}_g \Vert_2^2}}_{\text{(a)}} - \underbrace{\sum_{g'=1,{g'}\neq g}^{G} \frac{\tilde{\bh}_{g'} \left(\tilde{\bh}_{g'}^\mathrm{H}\tilde{\bh}_g\right)}{\Vert \tilde{\bh}_{g'}\Vert_2^2\Vert \tilde{\bh}_g\Vert_2^2}}_{\text{(b)}}\right). \label{approximated beamformer}
\end{align}
From \eqref{approximated beamformer}, we can find the two important observations. First, the beamforming vector $\bff_{{\mathrm{ZF}},g}$ can be represented with multiplications and summations of the RCs by using the NS expansion method as we intended. Second, the beamformer $\bff_{{\mathrm{ZF}},g}$ in \eqref{approximated beamformer} can be separated into the two parts where (a) corresponds to the signal part intended to the $g$-th group, and (b) is to null out the interference part. Specifically, (b) implies that the proposed MTZF beamforming is designed to reduce the influence from the other group's RCs on $\tilde{\bh}_g$ by subtracting the orthogonal projections of $\tilde{\bh}_g$. Consequently, some power, which is originally intended to the $g$-th group, is consumed. Based on these observations, we develop an effective multiple RISs reflection coefficients design that makes up for the loss of MTZF beamforming in the next subsection.

\subsection{Proposed Loss Minimizing Multiple RISs Design}

With \eqref{approximated beamformer}, it is possible to quantify the loss of MTZF beamforming at each user with channel vectors. For the $g_k$-th user, the intended signal loss can be represented by the inner product between the user's channel vector $\bh_{g_k}$ and (b) of $\bff_{{\mathrm{ZF}},g}$ in \eqref{approximated beamformer}. Then, the loss at the $g_k$-th user is given by
\begin{align}
	\sqrt{p_{\mathrm{ZF},g}}\bh_{g_k}^\mathrm{H} \left( \sum_{{g'}=1,{g'}\neq g}^{G} \frac{\tilde{\bh}_{g'} \left(\tilde{\bh}_{g'}^\mathrm{H}\tilde{\bh}_g\right)}{\Vert \tilde{\bh}_{g'}\Vert_2^2\Vert \tilde{\bh}_g\Vert_2^2} \right).
	\label{loss at each user}
\end{align}
In the multi-group multicasting systems, the beamforming vector $\bff_{{\mathrm{ZF}},g}$ and the RIS reflection coefficient matrix $\bPhi_g$ affect the same for all users in group $g$. Therefore, the total intended signal loss at group $g$ will be given by the summation of loss at each user in \eqref{loss at each user}, and we can design $\bPhi_g$ to minimize the total power of intended signal loss at group $g$. Assuming the other group's RIS reflection coefficient matrices are fixed, the optimization problem $(\mathrm{P}3)$ is given as
\begin{align}
	(\mathrm{P}3): &\min_{\bPhi_g}  \enspace \left\vert \sqrt{p_{\mathrm{ZF},g}}\sum_{k=1}^{K_g} \bh_{g_k}^\mathrm{H} \left( \sum_{{g'}=1,{g'}\neq g}^{G} \frac{\tilde{\bh}_{g'} \left(\tilde{\bh}_{g'}^\mathrm{H}\tilde{\bh}_g\right)}{\Vert \tilde{\bh}_{g'}\Vert_2^2\Vert \tilde{\bh}_g\Vert_2^2} \right)\right\vert.
\end{align}
Here, we can reformulate the objective function of $(\mathrm{P}3)$ with the definition of RCs in \eqref{rerpesentative channel definition} as
\begin{align}
	  & \left\vert \sqrt{p_{\mathrm{ZF},g}}\sum_{k=1}^{K_g} \bh_{g_k}^\mathrm{H} \left( \sum_{{g'}=1,{g'}\neq g}^{G} \frac{\tilde{\bh}_{g'} \left(\tilde{\bh}_{g'}^\mathrm{H}\tilde{\bh}_g\right)}{\Vert \tilde{\bh}_{g'}\Vert_2^2\Vert \tilde{\bh}_g\Vert_2^2} \right)\right\vert \notag                    \\
	= & \left\vert\sqrt{p_{\mathrm{ZF},g}} K_g \tilde{\bh}^\mathrm{H}_g \left( \sum_{{g'}=1,{g'}\neq g}^{G} \frac{\tilde{\bh}_{g'} \left(\tilde{\bh}_{g'}^\mathrm{H}\tilde{\bh}_g\right)}{\Vert \tilde{\bh}_{g'}\Vert_2^2\Vert \tilde{\bh}_g\Vert_2^2} \right)\right\vert, \notag                            \\
	= & \left\vert\sqrt{p_{\mathrm{ZF},g}} K_g \frac{\tilde{\bh}_g^\mathrm{H}\left( \textstyle\sum_{{g'}=1,{g'}\neq g}^{G}\frac{\tilde{\bh}_{g'}}{\Vert\tilde{\bh}_{g'}\Vert_2}\frac{\tilde{\bh}_{g'}^\mathrm{H}}{\Vert\tilde{\bh}_{g'}\Vert_2} \right)\tilde{\bh}_g}{\Vert \tilde{\bh}_g\Vert_2^2}\right\vert.
	\label{reformulated objective function}
\end{align}
For notational simplicity, we define
\begin{align}
	\bA_g \triangleq \sum_{{g'}=1,{g'}\neq g}^{G}\frac{\tilde{\bh}_{g'}}{\Vert\tilde{\bh}_{g'}\Vert_2}\frac{\tilde{\bh}_{g'}^\mathrm{H}}{\Vert\tilde{\bh}_{g'}\Vert_2},
	\label{A_g definition}
\end{align}
which is an $N \times N$ positive semi-definite matrix.\footnote{A rank-one matrix, which is the outer product of two identical column vectors, is positive semi-definite. Also, the summation of positive semi-definite matrices still preserves the property.} It is obvious that $\ba^\mathrm{H} \bA_g \ba \geq 0$ for any complex vector $\ba \in \mathbb{C}^{N \times 1}$, and the objective function in \eqref{reformulated objective function} becomes the real-valued function whose value is always greater than or equal to zero. This implies it is possible to remove the absolute value operation in $(\mathrm{P}3)$, which gives the reformulated problem $(\mathrm{P}3^\prime)$ as
\begin{align}
	(\mathrm{P}3^\prime): \min_{\bPhi_g}  \enspace	\frac{\tilde{\bh}_g^\mathrm{H}\bA_g\tilde{\bh}_g}{\Vert \tilde{\bh}_g\Vert_2^2}.
	\label{P1' definition}
\end{align}

The problem $(\mathrm{P}3^\prime)$; however, is still non-convex due to the unit-norm constraint on the main diagonal elements of $\bPhi_g$. For our system model in Section \ref{System Model}, only the single reflected signal through the RIS $g$ affects the $g$-th group users, and any variation on $\bPhi_g$ only affects $\tilde{\bh}_g$. Hence, to get insight into effective $\bPhi_g$ for $(\mathrm{P}3^\prime)$, we replace the optimization variable of $(\mathrm{P}3^\prime)$ with $\tilde{\bh}_g$, which gives another optimization problem $(\mathrm{P}4)$ as
\begin{align}
	(\mathrm{P}4): \min_{\tilde{\bh}_g}  \enspace 	\frac{\tilde{\bh}_g^\mathrm{H}\bA_g\tilde{\bh}_g}{\Vert \tilde{\bh}_g\Vert_2^2}.
\end{align}
Note that $(\mathrm{P}4)$ is a well-known Rayleigh quotient problem where the quotient reaches its minimum value, i.e., the minimum eigenvalue of the given Hermitian matrix $\bA_g$, when $\tilde{\bh}_g$ becomes the corresponding minimum eigenvector $\bv_{\mathrm{min},g}$.

Through the change of optimization variable, the formulation and optimal solution of $(\mathrm{P}4)$ suggest that the original objective function of $(\mathrm{P}3^\prime)$ can be minimized if the phase shifts of $\bPhi_g$ is designed to make $\tilde{\bh}_g$ close to $\bv_{\mathrm{min},g}$ as much as possible. For the two column vectors, the mathematical closeness is defined by the value of inner product. With this design purpose, we set the optimization problem $(\mathrm{P}5)$ with given $\bv_{\mathrm{min},g}$ as
\begin{align}
	(\mathrm{P}5): \max_{\bPhi_g} \enspace \left \vert \tilde{\bh}_g^\mathrm{H} \bv_{\mathrm{min},g}\right \vert.
	\label{P3 definition}
\end{align}
The objective function of $(\mathrm{P}5)$ can be expressed using the definition of RC in \eqref{rerpesentative channel definition} as
\begin{align}
	                           & \left \vert \left(\tilde{\bh}_{\mathrm{d},g}^\mathrm{H}+\tilde{\bh}_{\mathrm{r},g}^\mathrm{H}\bPhi_g \bH_g\right)\bv_{\mathrm{min},g}  \right \vert \notag                      \\
	= & \left \vert \left(\tilde{\bh}_{\mathrm{d},g}^\mathrm{H}+\bphi_g^\mathrm{H} \diag\left(\tilde{\bh}_{\mathrm{r},g}\right) \bH_g\right)\bv_{\mathrm{min},g}  \right \vert, \notag  \\
	=                          & \left \vert \tilde{\bh}_{\mathrm{d},g}^\mathrm{H}\bv_{\mathrm{min},g} +  \bphi_g^\mathrm{H} \diag\left(\tilde{\bh}_{\mathrm{r},g}\right) \bH_g\bv_{\mathrm{min},g}\right \vert.
	\label{P3 objective function}
\end{align}
Then, the $g$-th optimal RIS reflection coefficient vector $\bphi_{\mathrm{ZF},g}^\star$ that maximizes \eqref{P3 objective function} can be obtained as
\begin{align}
	\bphi_{\mathrm{ZF},g}^\star=\mathrm{exp}\left(j\angle\left(\bv_{\mathrm{min},g}^\mathrm{H}\tilde{\bh}_{\mathrm{d},g} \times \diag\left(\tilde{\bh}_{\mathrm{r},g}\right) \bH_g\bv_{\mathrm{min},g}\right)\right),
	\label{optimal RIS design}
\end{align}
where the optimality of closed-form solution in \eqref{optimal RIS design} can be proven by the following lemma.

\begin{lemma}
	For any complex number $\alpha$ and complex vector $\boldsymbol{\beta}$, the optimal solution of $\boldsymbol{\theta}$ that maximizes $\left\vert \alpha+\boldsymbol{\theta}^\mathrm{H} \boldsymbol{\beta} \right\vert $ with unit-norm constraint on the elements of $\boldsymbol{\theta}$ is given by $e^{j\angle(\alpha^*\times \boldsymbol{\beta})}$.
	\begin{proof}
		We can represent the square of the objective function $\left\vert \alpha+\boldsymbol{\theta}^\mathrm{H} \boldsymbol{\beta} \right\vert $ as
		\begin{align}
			&\left\vert \alpha+\boldsymbol{\theta}^\mathrm{H} \boldsymbol{\beta} \right\vert^2 \notag \\
			= & \left(\alpha^*+ \boldsymbol{\beta}^\mathrm{H} \boldsymbol{\theta} \right) \left(\alpha+ \boldsymbol{\theta}^\mathrm{H}\boldsymbol{\beta} \right), \notag \\
			= & \left( \alpha^* + \sum_n \beta^*(n) \theta(n) \right) \left( \alpha+\sum_n \theta^*(n) \beta(n)\right), \notag                                               \\
			= & \vert \alpha \vert^2+\sum_{n,m} \beta^*(n) \theta(n) \theta^*(m) \beta(m) \notag \\ &+ \sum_{n} \alpha^* \beta(n) \theta^*(n) + \sum_{n} \alpha \beta^*(n) \theta(n),\notag                      \\
			= & \vert \alpha \vert^2+\sum_{n} \vert \beta(n) \vert^2 +\sum_{n<m} 2 \mathrm{Re} (\beta^*(n)\theta(n) \theta^*(m) \beta(m)) \notag \\
			&+ \sum_{n} 2 \mathrm{Re} (\alpha^* \beta(n) \theta^*(n)).
			\label{proof of optimal solution}
		\end{align}
		Neglecting the constant terms, which are independent to $\boldsymbol{\theta}$, the real value $\mathrm{Re} (\alpha^* \beta(n) \theta^*(n))$ is maximized when $\theta(n)=e^{j\angle(\alpha^*\beta(n))}$ that also maximizes $\mathrm{Re} (\beta^*(n)\theta(n) \theta^*(m) \beta(m))$ with $\theta(m)=e^{j\angle(\alpha^*\beta(m))}$. This finishes the proof.
	\end{proof}
\end{lemma}

Note that, by its definition, $\bA_g$ is the summation of $G-1$ rank-one matrices, and the rank of $\bA_g$ cannot exceed $G-1$, i.e., $\mathrm{rank}(\bA_g) \leq G-1$. In practice, the BS would deploy a large number of antennas to serve many users simultaneously in the multi-group multicasting systems, and the number of groups $G$ would be smaller than the number of BS antennas $N$. Then, at least $N-(G-1)$ eigenvalues of $\bA_g$ will be zero, which is obviously the minimum since $\bA_g$ is the positive semi-definite matrix whose eigenvalues are greater than or equal to zero. Hence, we have multiple candidate pairs of $\bv_{\mathrm{min},g}$ and $\bphi_{\mathrm{ZF},g}^\star$. The best pair is the one that maximizes the objective function in \eqref{P3 objective function}. To obtain the best pair, we can first plug in $\bphi_{\mathrm{ZF},g}^\star$ in \eqref{optimal RIS design} into \eqref{P3 objective function}, which becomes
\begin{align}
	  & \left\vert \tilde{\bh}_{\mathrm{d},g}^\mathrm{H}\bv_{\mathrm{min},g} \right\vert + \left\Vert \diag\left(\tilde{\bh}_{\mathrm{r},g}\right) \bH_g\bv_{\mathrm{min},g} \right\Vert_1 \notag \\
	= & \left\Vert
	\begin{bmatrix}
		\tilde{\bh}_{\mathrm{d},g}^\mathrm{H} \\
		\diag\left(\tilde{\bh}_{\mathrm{r},g}\right) \bH_g
	\end{bmatrix}
	\bv_{\mathrm{min},g}\right\Vert_1.
	\label{pair criterion}
\end{align}
Among the candidates of $\bv_{\mathrm{min},g}$, we take the one that maximizes  \eqref{pair criterion}, which we denote as $\bv_{\mathrm{min},g}^\star$. Then, the final $\bphi_{\mathrm{ZF},g}^\star$ can be obtained as in \eqref{optimal RIS design} with $\bv_{\mathrm{min},g}^\star$. In this way, we need to perform \eqref{optimal RIS design} only once instead of deriving all possible candidates of $\bphi_{\mathrm{ZF},g}^\star$.

\alglanguage{pseudocode}
\begin{algorithm}[t]
	\caption{Proposed MTZF beamforming with loss minimizing multiple RISs design}
	\textbf{Initialize}
	\begin{algorithmic}[1]
		\State Set $\epsilon_\mathrm{norm} >0$
		\State Initialize $\{\bphi_{\mathrm{ZF},g}\}_{g=1}^G$
		\State Calculate $\{\tilde{\bh}_g\}_{g=1}^G$ by \eqref{rerpesentative channel definition}
		\State Compute $T_\mathrm{tmp}=\sum_{g=1}^G \Vert \bphi_{\mathrm{ZF},g}\Vert_2$
	\end{algorithmic}
	\textbf{Iterative update}
	\begin{algorithmic}[1]
		\addtocounter{ALG@line}{+4}
		\For{$i_2=1,\cdots,I_2$}
		\For{$g=1,\cdots,G$}
		\State Calculate $\bA_g$ by \eqref{A_g definition}
		\State Update candidate set of $\bv_{\mathrm{min},g}$
		\State Update $\bv_{\mathrm{min},g}^\star$ that maximizes \eqref{pair criterion}
		\State Update $\bphi_{\mathrm{ZF},g}^\star$ by \eqref{optimal RIS design} with $\bv_{\mathrm{min},g}^\star$
		\EndFor
		\State Compute $T=\sum_{g=1}^{G}\left \Vert \bphi_{\mathrm{ZF},g}^\star \right \Vert_2$
		\If {$\vert T-T_\mathrm{tmp}\vert < \epsilon_\mathrm{norm}$}
		\State Break
		\Else
		\State Set $T_\mathrm{tmp}=T$
		\EndIf
		\EndFor
		\State Calculate $\{\tilde{\bh}_g\}_{g=1}^G$ by \eqref{rerpesentative channel definition} with $\{\bphi_{\mathrm{ZF},g}^\star\}_{g=1}^G$
		\State Calculate $\bF_\mathrm{ZF}$ by \eqref{MTZF beamforming}
	\end{algorithmic}
\end{algorithm} 

\subsection{Algorithm Description and Complexity Analysis}

The overall process of the proposed MTZF beamforming with the multiple RISs design that minimizes the loss of MTZF beamforming is described in Algorithm 2. We first want to emphasize that the solution in \eqref{optimal RIS design} does not depend on the actual values of the MTZF beamforming vectors $\{\bff_{{\mathrm{ZF}},g}\}_{g=1}^G$ and only exploits the inherent characteristics of the beamformer. Therefore, there is no AO between $\{\bphi_{\mathrm{ZF},g}\}_{g=1}^G$ and $\{\bff_{{\mathrm{ZF}},g}\}_{g=1}^G$ different from the most of RIS-assisted communication techniques including the proposed technique in Section \ref{proposed BD+min. rate maximization RIS} \cite{RIS_intro_5,RIS_multicasting_1(single),Multi_RIS_1}. The effective RIS reflection coefficients are first updated in an iterative way, and the update can stop when the iteration index $i_2$ reaches its maximum value $I_2$ or the total change of RIS reflection coefficients becomes less than the threshold $\epsilon_\mathrm{norm}$. With the designed RIS reflection coefficient vectors $\{\bphi_{\mathrm{ZF},g}^\star\}_{g=1}^G$, the final RCs can be obtained with the definition in \eqref{rerpesentative channel definition}. At last, the MTZF beamforming matrix $\bF_\mathrm{ZF}$ is acquired with the final RCs as in \eqref{MTZF beamforming}.

We highlight that our proposed Algorithm 2 can be conducted with low computational complexity and is practical for the two following reasons; 1) there is no AO, and 2) the closed-form solution in \eqref{optimal RIS design} only requires basic linear operations. Assuming $N \gg G$, the complexity per iteration of the multiple RISs reflection coefficients design is given by $\cO\big(\sum_{g=1}^G\left(N^3+N^2M_g\right)\big)$, which is mainly dominated by the SVD of $\bA_g$ \cite{multicasting_block_diag}. Also, the complexity associated with the final MTZF beamforming is $\cO\big(\sum_{g=1}^G(NM_g)+NG^2\big)$, and therefore, the maximum overall complexity becomes $\cO\big(I_2 \sum_{g=1}^G\big(N^3+N^2M_g\big)+ \sum_{g=1}^G (NM_g)+NG^2\big)$. Since it does not depend on the total number of users $K$ and only grows linearly with the number of RIS elements, the BS can simultaneously serve many users with the aid of large scale RISs efficiently.


\section{Numerical Results} \label{numerical results}

\begin{figure}[t]
	\centering
	\includegraphics[width=1.05\columnwidth]{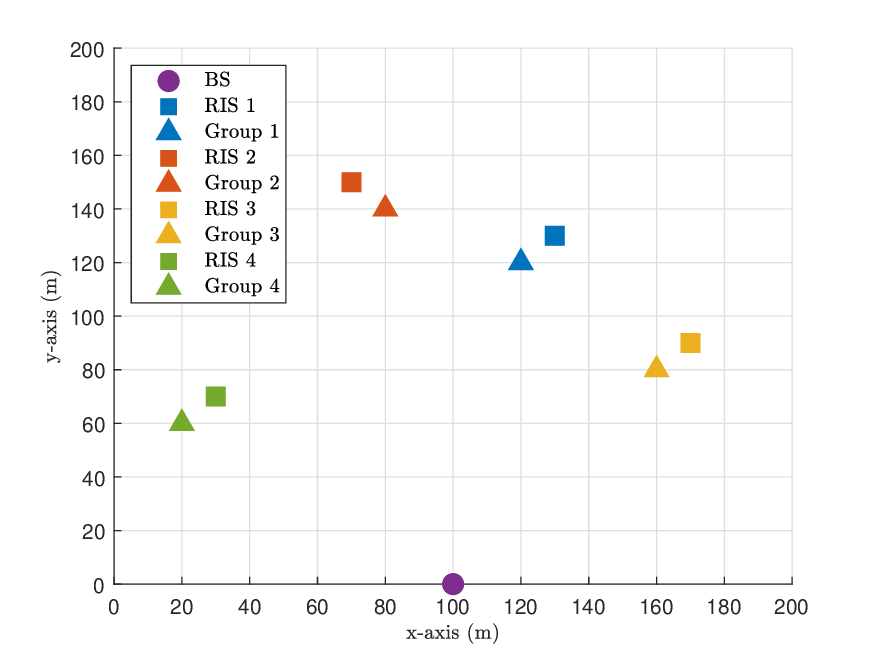}
	\caption{Positions of BS and each group's center with distributed RISs when $G=4$.}
	\label{location_fig}
\end{figure}

In this section, we evaluate the performance of proposed techniques in Sections \ref{proposed BD+min. rate maximization RIS} and \ref{proposed MTZF+loss minimization RIS} for the multi-group multicasting systems. Taking a three-dimensional (3D) coordinate system into account, Fig.~\ref{location_fig} represents the positions of BS and each group's center with the distributed RISs\footnote{Note that the deployments of multiple RISs will affect the overall system performance, and high channel correlation due to closely located RISs can cause noticeable performance degradation. Therefore, it is desirable to deploy RISs far apart as long as they can support all groups.} in the xy-plane when $G=4$. In each group, single-antenna users are uniformly distributed in a circle area with the radius of 5 m around each center point. The height of BS, RISs, and users are set as 15 m, 5 m, and 1 m, respectively. For practical setups, the BS antennas and elements of each RIS are deployed in uniform planar array structures assuming half-wavelength antenna and element spacing where the arrays are aligned parallel to the xz-plane. We consider $N_\mathrm{ver}$ vertical and $N_\mathrm{hor}$ horizontal antennas for the BS where $N=N_\mathrm{ver}\times N_\mathrm{hor}$. Similarly, for the $g$-th RIS, $M_{\mathrm{ver},g}$ vertical and $M_{\mathrm{hor},g}$ horizontal elements are deployed where $M_g=M_{\mathrm{ver},g} \times M_{\mathrm{hor},g}$. The noise variance at each user is set as $\sigma_{g_k}^2 =-114$~dBm.

We adopt the Rician fading model consists of one line-of-sight (LoS) path and multiple non-line-of-sight (NLoS) paths for all channel links as in \cite{RIS_intro_3,Rician_channel}. For example, the downlink channel between the BS and RIS $g$, $\bH_g$, is 
\begin{align}
	&\bH_g=\sqrt{\mu_0\left(d_{\mathrm{BR},g}/d_0\right)^{-\eta_{\mathrm{BR}}}} \sqrt{\frac{{M_g N}}{1+\kappa_\mathrm{BR}}} \notag \\
	&\times \bigg(\sqrt{\kappa_\mathrm{BR}}\gamma_{\mathrm{BR},g}^{0} \ba_{\mathrm{RIS},g} \left( \nu_{\mathrm{BR},g}^{\mathrm{rx},0},\xi_{\mathrm{BR},g}^{\mathrm{rx},0} \right)\ba_{\mathrm{BS}}^\mathrm{H} \left( \nu_{\mathrm{BR},g}^{\mathrm{tx},0},\xi_{\mathrm{BR},g}^{\mathrm{tx},0} \right) \notag \\
	&+ \frac{1}{\sqrt{C_{\mathrm{BR}}}} \sum_{c=1}^{C_{\mathrm{BR}}}\gamma_{\mathrm{BR},g}^{c}\ba_{\mathrm{RIS},g} \left( \nu_{\mathrm{BR},g}^{\mathrm{rx},c},\xi_{\mathrm{BR},g}^{\mathrm{rx},c} \right) \notag \\
	& \times \ba_{\mathrm{BS}}^\mathrm{H} \left( \nu_{\mathrm{BR},g}^{\mathrm{tx},c},\xi_{\mathrm{BR},g}^{\mathrm{tx},c} \right) \bigg), \label{Rician fading channel model}
\end{align}
where we set the path loss $\mu_0=-30$ dB at the reference distance $d_0=1$ m, and the distance between the BS and RIS~$g$ is denoted by $d_{\mathrm{BR},g}$. The path loss exponent, Rician factor, and number of NLoS paths between the BS and each RIS are denoted by $\eta_\mathrm{BR}$, $\kappa_\mathrm{BR}$, and $C_{\mathrm{BR}}$, respectively. For the $c$-th path, $\gamma_{\mathrm{BR},g}^{c} \sim \cC\cN (0,1)$ is the complex gain, and we denote the vertical and horizontal arrival angles at the RIS~$g$ as $\nu_{\mathrm{BR},g}^{\mathrm{rx},c}$ and $\xi_{\mathrm{BR},g}^{\mathrm{rx},c}$. Similarly, we denote the vertical and horizontal departure angles at the BS as $\nu_{\mathrm{BR},g}^{\mathrm{tx},c}$ and $\xi_{\mathrm{BR},g}^{\mathrm{tx},c}$. With the half wavelength spacing assumption, the array response vector at the $g$-th RIS, $\ba_{\mathrm{RIS},g}(\cdot)$, is given~as
\begin{align}
	&\ba_{\mathrm{RIS},g} \left( \nu_{\mathrm{BR},g}^{\mathrm{rx},c},\xi_{\mathrm{BR},g}^{\mathrm{rx},c} \right) \notag \\
	=& \frac{1}{\sqrt{M_g}} \left[1,\cdots,e^{j\pi\left(M_{\mathrm{ver},g}-1\right)\sin\left(\nu_{\mathrm{BR},g}^{\mathrm{rx},c}\right)} \right]^\mathrm{T} \notag \\
	&\otimes \left[1,\cdots,e^{j\pi\left(M_{\mathrm{hor},g}-1\right)\sin\left(\xi_{\mathrm{BR},g}^{\mathrm{rx},c}\right)\cos\left(\nu_{\mathrm{BR},g}^{\mathrm{rx},c}\right)} \right]^\mathrm{T},
\end{align} 
and the array response vector at the BS,  $\ba_{\mathrm{BS}}(\cdot)$, is given as
\begin{align}
	&\ba_{\mathrm{BS}} \left( \nu_{\mathrm{BR},g}^{\mathrm{tx},c},\xi_{\mathrm{BR},g}^{\mathrm{tx},c} \right) \notag \\
	=& \frac{1}{\sqrt{N}} \left[1,\cdots,e^{j\pi\left(N_{\mathrm{ver}}-1\right)\sin\left(\nu_{\mathrm{BR},g}^{\mathrm{tx},c}\right)} \right]^\mathrm{T} \notag \\
	&\otimes \left[1,\cdots,e^{j\pi\left(N_{\mathrm{hor}}-1\right)\sin\left(\xi_{\mathrm{BR},g}^{\mathrm{tx},c}\right)\cos\left(\nu_{\mathrm{BR},g}^{\mathrm{tx},c}\right)} \right]^\mathrm{T}.
\end{align} 

\begin{table}[t]
	\renewcommand{\arraystretch}{2} 
	\centering
	\caption{Rician fading channel parameters.}
	\begin{tabular}{| c | c | c | c |}
		\hline
		\multirow{2}*{Communication links} & \multirow{2}*{\shortstack{Path loss                                          \\exponents}} & \multirow{2}*{\shortstack{Rician \\ factors}} & \multirow{2}*{\shortstack{Number of\\NLoS paths}} \\ & & & \\
		\hline
		\multirow{1}*{BS-user direct}      & \multirow{1}*{$\eta_{\mathrm{BU}}=4.5$}                  & \multirow{1}*{$\kappa_{\mathrm{BU}}=3 \text{ }\mathrm{dB}$} & \multirow{1}*{$C_\mathrm{BU}=8$} \\
		\hline
		\multirow{1}*{RIS-user reflection} & \multirow{1}*{$\eta_{\mathrm{RU}}=2.2$}                  & \multirow{1}*{$\kappa_{\mathrm{RU}}=7 \text{ }\mathrm{dB}$} & \multirow{1}*{$C_\mathrm{RU}=4$} \\
		\hline
		\multirow{1}*{BS-RIS}              & \multirow{1}*{$\eta_{\mathrm{BR}}=2.3$}                  & \multirow{1}*{$\kappa_{\mathrm{BR}}=5 \text{ }\mathrm{dB}$} & \multirow{1}*{$C_\mathrm{BR}=8$} \\
		\hline
	\end{tabular}
	\label{Table. channel parameter}
\end{table}

Note that $\bh_{\mathrm{d},g_k}^\mathrm{H}$ and $\bh_{\mathrm{r},g_k}^\mathrm{H}$ are defined similarly as in \eqref{Rician fading channel model} with proper modifications on the distance, path loss exponent, Rician factor, number of NLoS paths, angles, and array response vectors. The specific values of remaining parameters that construct the Rician fading channels are given in Table~I. For all channel links, the arrival and departure angles of NLoS paths are randomly generated within $5^\circ$ for vertical and $8^\circ$ for horizontal angular spread centered at the LoS paths, which are numerically calculated with the actual positions of transceivers. To reveal the homogeneity among the groups, we assume the same number of users and RIS elements for each group, i.e., $K_1=\cdots=K_G$ and $M_1=\cdots=M_G$. 

As a baseline, we consider the design in \cite{RIS_multicasting_1(single)}, which directly targets the maximization of the sum-rate of all groups in \eqref{multicasting sum rate} using only a single RIS. For a fair comparison, we assume the baseline exploits one RIS that deploys $M$ elements, which is the total number of elements for all distributed multiple RISs, i.e., $M=\sum_{g=1}^G M_g$. By exploiting the majorization-minimization algorithm framework, the design in \cite{RIS_multicasting_1(single)} also adopts the AO method to optimize the transmit beamformer including the power allocation at the BS and the single-RIS reflection coefficients. For each iteration, the optimization problem is reformulated as the second-order cone programming problem where the solution can be obtained by using the existing solvers like CVX \cite{cvx}. Note that the computational complexity per iteration is given by $\cO(N^3K^3+NK^{4.5}+N^3K^{5.5}+K^{3.5}M+K^{2.5}M^3+M^{3.5})$, which is much higher than that of our proposed techniques, and the complexity increases dramatically with the total number of users $K$ and RIS elements $M$. For the numerical results, the single RIS is assumed to be deployed at [100,100,5] m for theoretical evaluation even though the physical reflection of signals transmitted from the BS to some users is infeasible. In addition, we compare the simple random cases where all phase shifts of the multiple RISs reflection coefficients are uniformly and randomly distributed in the range of $[0, 2\pi)$, and the BD-based or  MTZF beamforming is conducted.

\begin{figure}[t]
	\centering
	\includegraphics[width=1.05\columnwidth]{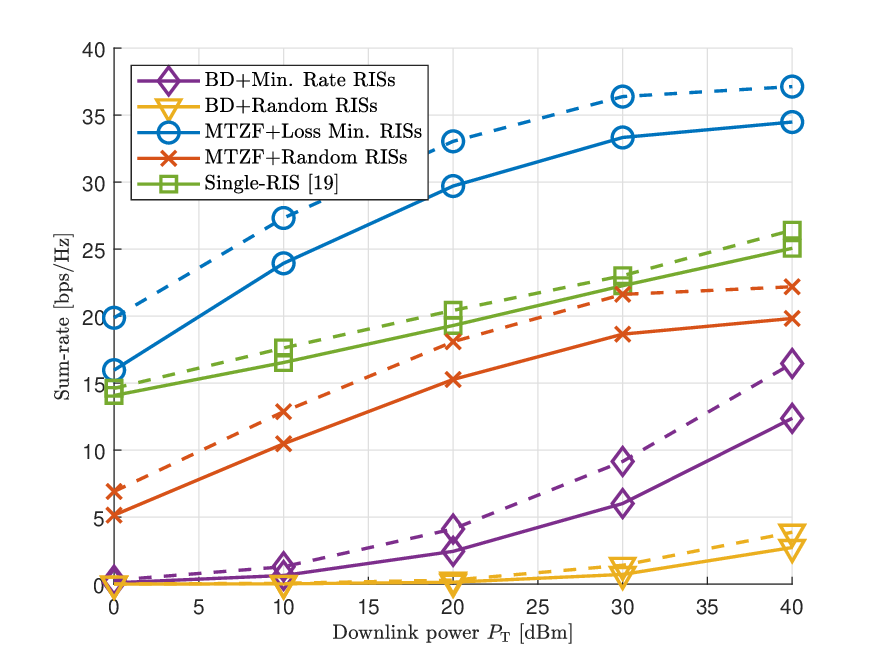}
	\caption{Average sum-rate performance according to $P_\mathrm{T}$ with $N=2 \times 8$, $G=3$, and $K_g=4$. The solid lines are the case of $M_g=8\times 3$, and $M_g=8\times 4$ is considered for the dashed lines.}
	\label{sum_rate_vs_downlink_power_RIS_elements}
\end{figure}

\begin{figure}[t]
	\centering
	\includegraphics[width=1.05\columnwidth]{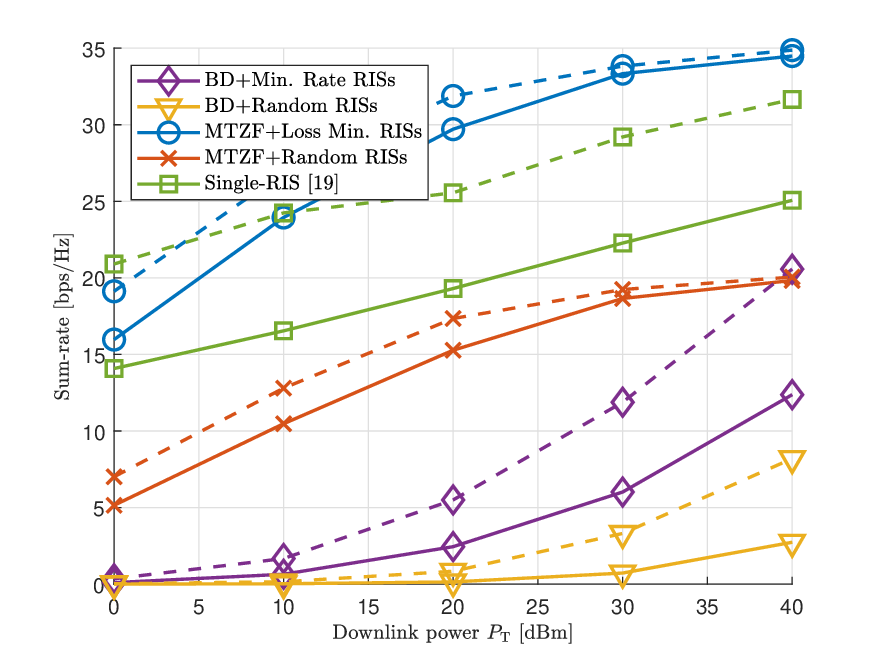}
	\caption{Average sum-rate performance according to $P_\mathrm{T}$ with $M_g=8\times 3$, $G=3$, and $K_g=4$. The solid lines are the case of $N=2\times 8$, and $N=3\times 8$ is considered for the dashed lines.}
	\label{sum_rate_vs_downlink_power_BS_antennas}
\end{figure}

Fig. \ref{sum_rate_vs_downlink_power_RIS_elements} shows the average sum-rate performance of the proposed techniques and baseline schemes according to the total transmit power $P_\mathrm{T}$ with different numbers of RIS elements $M_g$. The solid lines and dashed lines are the cases of $M_g=8\times 3$ and $M_g=8\times 4$, respectively. It can be observed that the MTZF beamforming-based approach outperforms the other schemes for all ranges of $P_\mathrm{T}$. 
Although it requires higher computational complexity, the BD-based approach shows poor sum-rate performance. The BD-based approach is able to eliminate the inter-group interference completely by exploiting the null-space projections. Without any inter-group interference, we can expect continuously increasing sum-rate as $P_\mathrm{T}$ increases. However, the technique has a fundamental weakness that it severely suffers from the loss of intended signal power from the interference cancellation considering all channels in the system, which results in low data rate for each user and therefore low sum-rate. In contrast, the MTZF beamforming-based approach efficiently mitigates the inter-group interference based on the RCs since the number of RCs is much less than the number of total channels, and the multiple RISs are carefully designed to minimize the loss of intended signal power. This makes the notable difference between the two proposed techniques in terms of the sum-rate performance. In addition, even with much lower computational complexity, the MTZF beamforming-based approach shows higher sum-rate performance than the baseline single-RIS system, and this entails the great effectiveness of deploying distributed RISs with their group-specific operations. As $M_g$ increases, the performance gain is more prominent for the proposed techniques by fully exploiting the increased element number of each distributed RIS, and the difference from the random RISs cases implies that it is essential to properly design the reflection coefficients in order to utilize the advantages of deploying multiple RISs.

\begin{figure}[t]
	\centering
	\includegraphics[width=1.05\columnwidth]{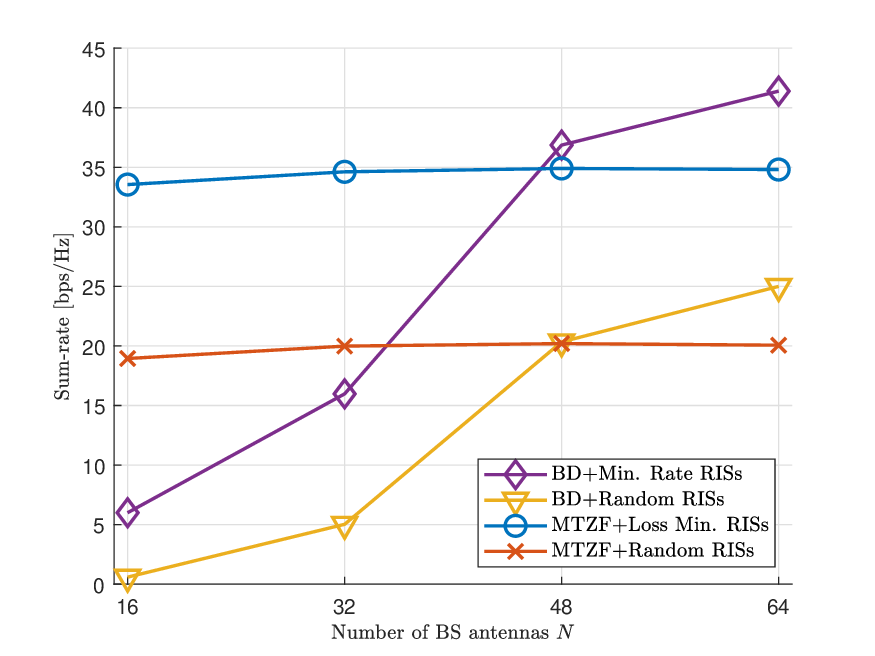}
	\caption{Average sum-rate performance according to $N$ with $M_g=8\times 3$, $G=3$, $K_g=4$, and $P_\mathrm{T}=$30 dBm.}
	\label{sum_rate_vs_BS_antennas}
\end{figure}

Similar to Fig. \ref{sum_rate_vs_downlink_power_RIS_elements}, we evaluate the average sum-rate results according to $P_\mathrm{T}$ in Fig. \ref{sum_rate_vs_downlink_power_BS_antennas} with different numbers of BS antennas $N$. The solid lines and dashed lines are the cases of $N=2\times 8$ and $N=3\times 8$, respectively. The overall trends among the proposed techniques and baselines are comparable to the results in Fig.~\ref{sum_rate_vs_downlink_power_RIS_elements}, and still the MTZF beamforming-based approach shows the highest sum-rate performance. However, there is the noticeable performance improvement for the BD-based approach especially in high $P_\mathrm{T}$ regime. As $N$ increases, it becomes much easier for the BD-based beamforming technique to eliminate the inter-group interference, and this implies that we can anticipate the reduced intended signal power loss and the increased sum-rate without any inter-group interference. This advantage of the BD-based approach with large $N$ becomes clearer in Fig. \ref{sum_rate_vs_BS_antennas} where the figure shows the average sum-rate performance according to $N$. Even though the number of $N$ increases, there is no remarkable performance improvement for the MTZF beamforming-based approaches. For the schemes that exploit the BD-based beamforming, however, the sum-rate increases dramatically with $N$, and we can expect the BD-based approach will show better sum-rate performance when the BS deploys a large dimensional antenna array.\footnote{For a large number of $N$, we numerically proved that the proposed BD-based approach shows similar performance to the case when the reflection coefficients of multiple RISs are obtained by the exhaustive search to maximize the minimum rate for each group assuming the BD-based beamforming.}

\begin{figure}[t]
	\centering
	\includegraphics[width=1.05\columnwidth]{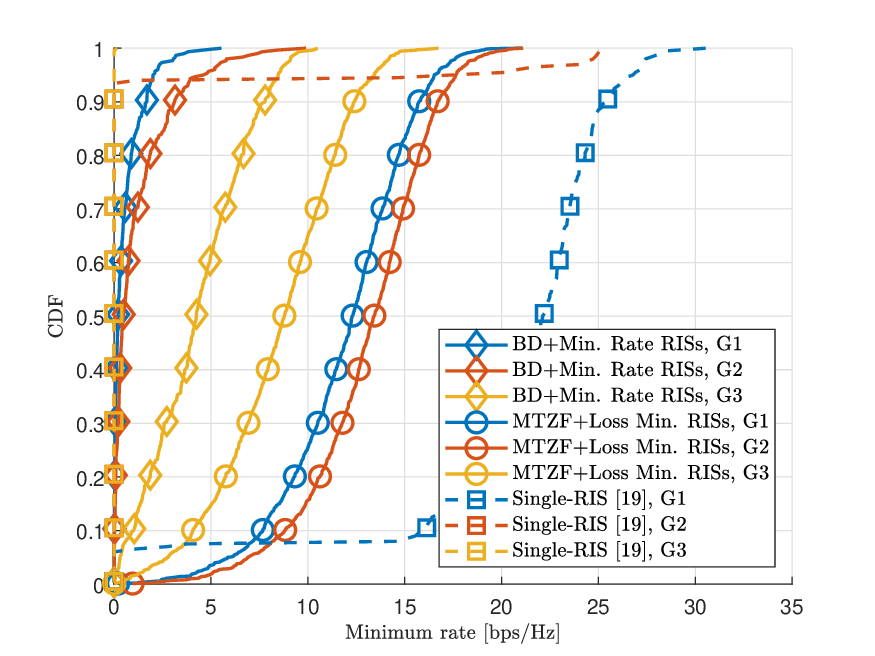}
	\caption{CDF of minimum rate when $N=2 \times 8$, $M_g=8\times 3$, $G=3$, $K_g=4$, and $P_\mathrm{T}=$30 dBm.}
	\label{CDF_vs_min_rate}
\end{figure}

As we can observe in Fig. \ref{sum_rate_vs_downlink_power_BS_antennas}, the single-RIS system outperforms the proposed techniques in low $P_\mathrm{T}$ regime. However, the single-RIS system has a critical weak point in terms of QoS for each group. 
In Fig. \ref{CDF_vs_min_rate}, the cumulative distribution function (CDF) of the minimum rate for each group is depicted. The single-RIS system alternately optimizes the beamformer at the BS and reflection coefficients at the RIS to maximize the sum-rate of all groups. Then, the result of optimization is inclined to support only one favored group to totally eliminate the inter-group interference and achieve high sum-rate. In the scenario of interest, the first group, G1, whose position is the closest to the RIS, is selected, and other groups have almost zero rate as clearly shown in Fig. \ref{CDF_vs_min_rate}. In contrast, the proposed techniques effectively mitigate the inter-group interference in their own way while they transmit independent data for each group with non-zero rates. Therefore, even though the single-RIS system achieves higher sum-rate than the proposed techniques in low $P_\mathrm{T}$ regime, the single-RIS system may be infeasible in practice because of the QoS constraints, e.g., the minimum rate requirement at each group. In addition, the results in Figs. \ref{sum_rate_vs_downlink_power_RIS_elements} to~\ref{CDF_vs_min_rate} clearly show that deploying multiple RISs with their group specific operations can bring performance improvements, and there is a fundamental limitation for the multi-group multicasting system exploiting a single RIS.

\begin{figure}[t]
	\centering
	\includegraphics[width=1.05\columnwidth]{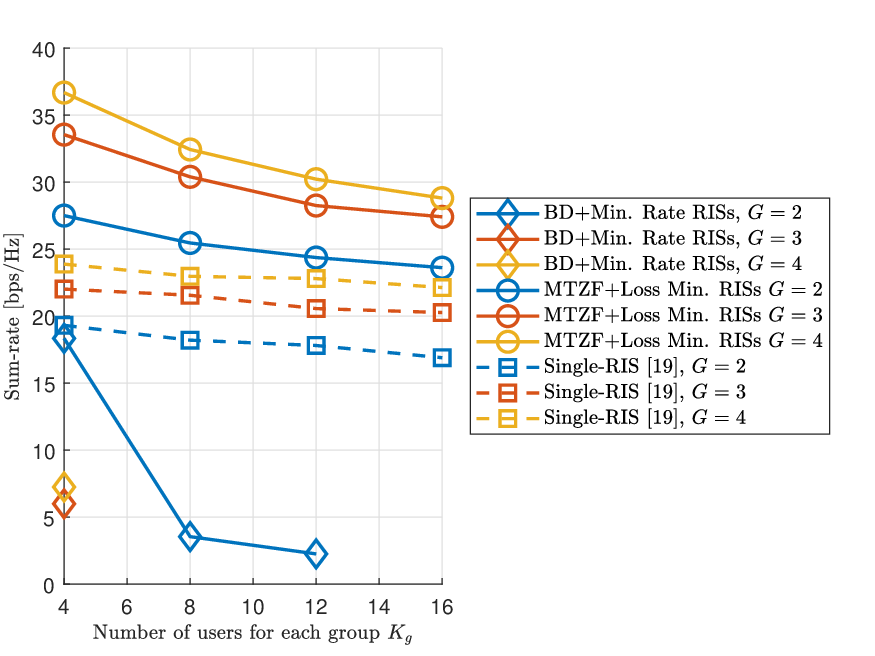}
	\caption{Average sum-rate performance according to $K_g$ when $N=2 \times 8$, $M_g=8\times 3$, and $P_\mathrm{T}=$30 dBm.}
	\label{sum_rate_vs_number_of_user}
\end{figure}

\begin{figure}[t]
	\centering
	\subfloat[Positions in xy-plane.]{
		\label{location_fig new loc}	
		\includegraphics[width=1.05\columnwidth]{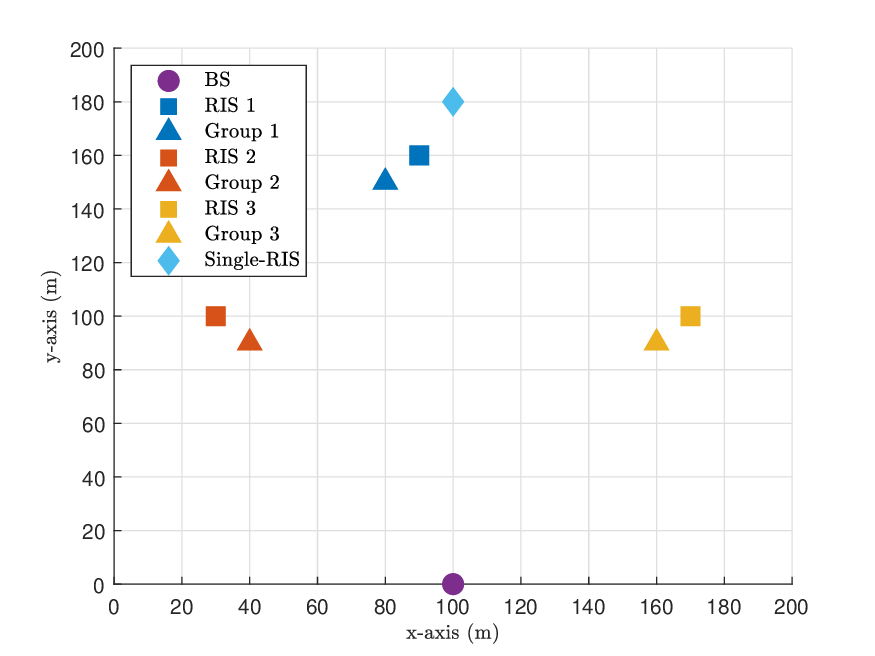}}
	\hfill
	\centering
	\subfloat[Average sum-rate.]{
		\label{sum_rate_vs_downlink_power_RIS_elements new loc}
	\includegraphics[width=1.05\columnwidth]{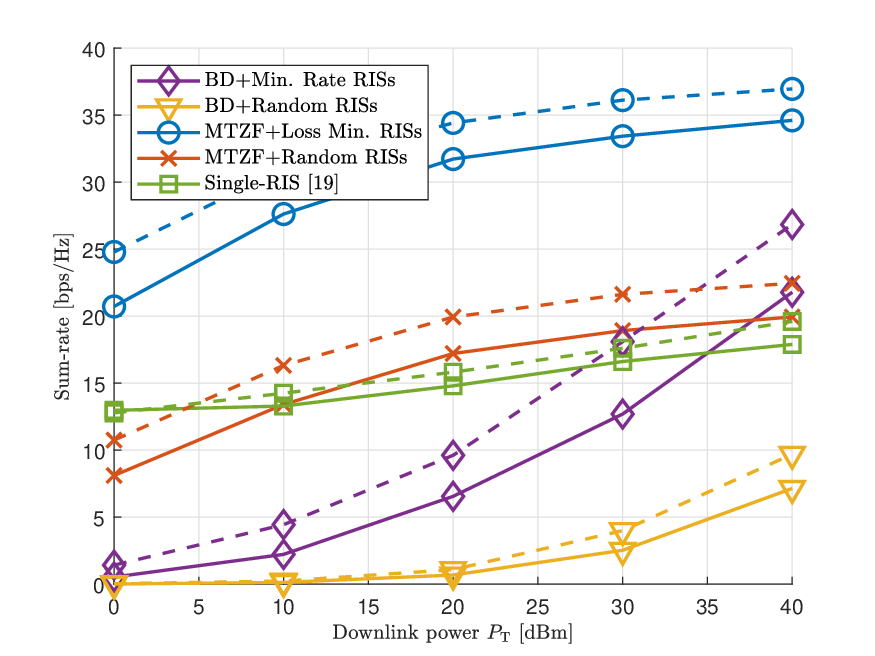}}
	\caption{Average sum-rate performance according to $P_\mathrm{T}$ when $N=2 \times 8$, $G=3$, and $K_g=4$ assuming different positions. The solid lines are the case of $M_g=8\times 3$, and $M_g=8\times 4$ is considered for the dashed lines.}
\end{figure}

In Fig. \ref{sum_rate_vs_number_of_user}, the sum-rate performance according to $K_g$ is depicted with different numbers of $G$. For example, $G=3$ means only the first three groups in Fig. \ref{location_fig} are considered for simulations. Note that the possible number of $K_g$ is limited for the BD-based beamforming technique when $N$ is fixed as we discussed in Section \ref{BD-based beamforming technique}; therefore, only a few results are derived for the BD-based approach. Basically, for all cases, the sum-rate decreases as $K_g$ increases. This is because the data rate of each group is limited by that of the worst case user following the multicasting nature, and the chance of some users having bad channel condition increases with large $K_g$. Still, the MTZF beamforming-based approach outperforms the single-RIS system by fully exploiting the multiple distributed RISs and supporting all groups with non-zero rates similar to the previous results. 

To evaluate the performance of proposed techniques with baseline schemes in a more practical environment where the single-RIS system can properly reflect incoming signals, we consider a different position setup for the BS and each group's center with the RISs when $G=3$ as in Fig. \ref{location_fig new loc}. Except for the positions on the xy-plane, all numerical parameters are the same as before. For the single-RIS system, the RIS is deployed at the outer side to enable the physical reflections through the single-RIS to all users. In Fig. \ref{sum_rate_vs_downlink_power_RIS_elements new loc}, we investigate the sum-rate performance as in Fig. \ref{sum_rate_vs_downlink_power_RIS_elements}. Similar to the results in Fig.~\ref{sum_rate_vs_downlink_power_RIS_elements}, the MTZF beamforming-based approach shows noticeably higher sum-rate than the single-RIS system for all ranges of $P_\mathrm{T}$. Also, the BD-based approach outperforms the single-RIS system in high $P_\mathrm{T}$ regime. This implies that it is difficult to assist the BS using only one RIS when the groups are apart from each other, and the multiple distributed RISs should be exploited with the proper reflection coefficients for multi-group multicasting~systems.

\section{Conclusion} \label{conclusion}

We proposed the two techniques for practical utilization of multiple RISs for the multi-group multicasting systems. As the first approach, the BD-based beamforming was adopted to eliminate the inter-group interference. Then, the multiple RISs were designed to maximize the minimum rate of each group following the multicasting nature. In addition, to mitigate the inter-group interference efficiently, the MTZF beamforming was developed by reformulating the linear ZF beamforming. For the multiple RISs, the reflection coefficients are designed to make up for the inevitable loss of MTZF beamforming. The effective closed-form solution for the RIS coefficients that minimize the loss was obtained independently to the actual values of MTZF beamformer and only requires basic linear operations, which makes the proposed design highly practical. 
Numerical results showed that the MTZF beamforming-based approach outperforms the baseline schemes in terms of the sum-rate and the minimum rate. Without any inter-group interference, the BD-based approach demonstrated its ability for achieving higher sum-rate when the number of antennas at the BS increases. Also, the results proved advantages of deploying multiple distributed RISs and applying appropriate group-specific RIS operations for the multi-group multicasting systems.

Possible future research directions can include practical channel estimation for the multi-group multicasting systems supported by multiple RISs. Although we exploit the instantaneous CSI for the proposed techniques, algorithm development based on the statistical CSI as in \cite{RIS_S_CSI_1,RIS_S_CSI_2} can be a promising solution to tackle the channel estimation overhead issue. To overcome the double path-loss effect and bring additional performance improvements, deploying active RISs as in \cite{active_RIS_1, active_RIS_2} can be adopted, and the extension of the proposed reflection coefficients designs for active RISs can be another interesting research topic. It is also worth investigating the extension of the proposed techniques to the joint communication and multi-target sensing system as in \cite{RIS_sensing} by fully exploiting the interference mitigation capabilities of proposed techniques. In addition, a joint approach of multicasting and unicast data transmissions can be considered to maximize the sum-rate or energy efficiency; still having the low computational complexity.

\bibliographystyle{IEEEtran}
\bibliography{Practical_Multi_RIS}

\end{document}